\documentclass[usenatbib,useAMS]{mn2e}
\usepackage{
graphicx,epstopdf,float,color,natbib,hyperref,amsmath,amssymb,deluxetable, pdflscape,multirow,psfig}
\usepackage{epstopdf}
\usepackage{bm}

\newcommand{\ba}{\begin{eqnarray}}
\newcommand{\ea}{\end{eqnarray}}
\newcommand{\be}{\begin{equation}}
\newcommand{\ee}{\end{equation}}
\renewcommand{\vec}{\bm}

\title[Cosmology with Doppler lensing]{Cosmology with Doppler lensing}
\author[D. J. Bacon et al.]{David J. Bacon$^1$, Sambatra Andrianomena$^2$, Chris Clarkson$^2$, 
\newauthor Krzysztof Bolejko$^3$, Roy Maartens$^{1,4}$ \\
$^1$Institute of Cosmology \& Gravitation, University of Portsmouth, Dennis Sciama Building, Portsmouth, PO1 3FX\\
$^2$Astrophysics, Cosmology Gravity Centre, and Department of Mathematics and Applied Mathematics, \\University of Cape Town, Cape Town, 7701, South Africa.\\
$^3$Sydney Institute for Astronomy,
The University of Sydney, NSW 2006, Australia\\
$^4$Physics Department, University of the Western Cape, 
Cape Town 7535, South Africa
}

\begin{document}
\date{Accepted ----. Received ----; in original form ----.}
\pagerange{\pageref{firstpage}--\pageref{lastpage}} \pubyear{2009}
\maketitle
\label{firstpage}

\begin{abstract}
Doppler lensing is the apparent change in object size and magnitude due to peculiar velocities. Objects falling into an overdensity appear larger on its near side, and smaller on its far side, than typical objects at the same redshifts. This effect dominates over the usual gravitational lensing magnification at low redshift. Doppler lensing is a promising new probe of cosmology, and we explore in detail how to utilize the effect with forthcoming surveys. We present cosmological simulations of the Doppler and gravitational lensing effects based on the Millennium simulation. We show that Doppler lensing can be detected around stacked voids or unvirialised over-densities. New power spectra and correlation functions are proposed which are designed to be sensitive to Doppler lensing. We consider the impact of gravitational lensing and intrinsic size correlations on these quantities. We compute the correlation functions and forecast the errors for realistic forthcoming surveys, providing predictions for constraints on cosmological parameters. Finally, we demonstrate how we can make 3-D potential maps of large volumes of the Universe using Doppler lensing.  
\end{abstract}

\section{Introduction}

Light rays from distant sources are focused by overdensities (or defocused by underdensities) along the line of sight, leading to apparent magnification (or demagnification) of images.  
But besides this {\em gravitational lensing}, there is a further effect which appears to magnify or demagnify the images of objects in the Universe. This {\em Doppler lensing} effect arises from the peculiar velocity of the source, and was first highlighted and investigated in general by \citet{2008PhRvD..78l3530B} (see also \citet{2006PhRvD..73b3523B}). \citet{2013PhRvL.110b1302B} then showed that the effect can dominate over  gravitational lensing, and even reverse its effect, leading to an `anti-lensing' phenomenon. Doppler lensing gives a new window into the peculiar velocity field in addition to the usual redshift space distortion measurements. 
\begin{figure}
\centering
\includegraphics[width=8cm]{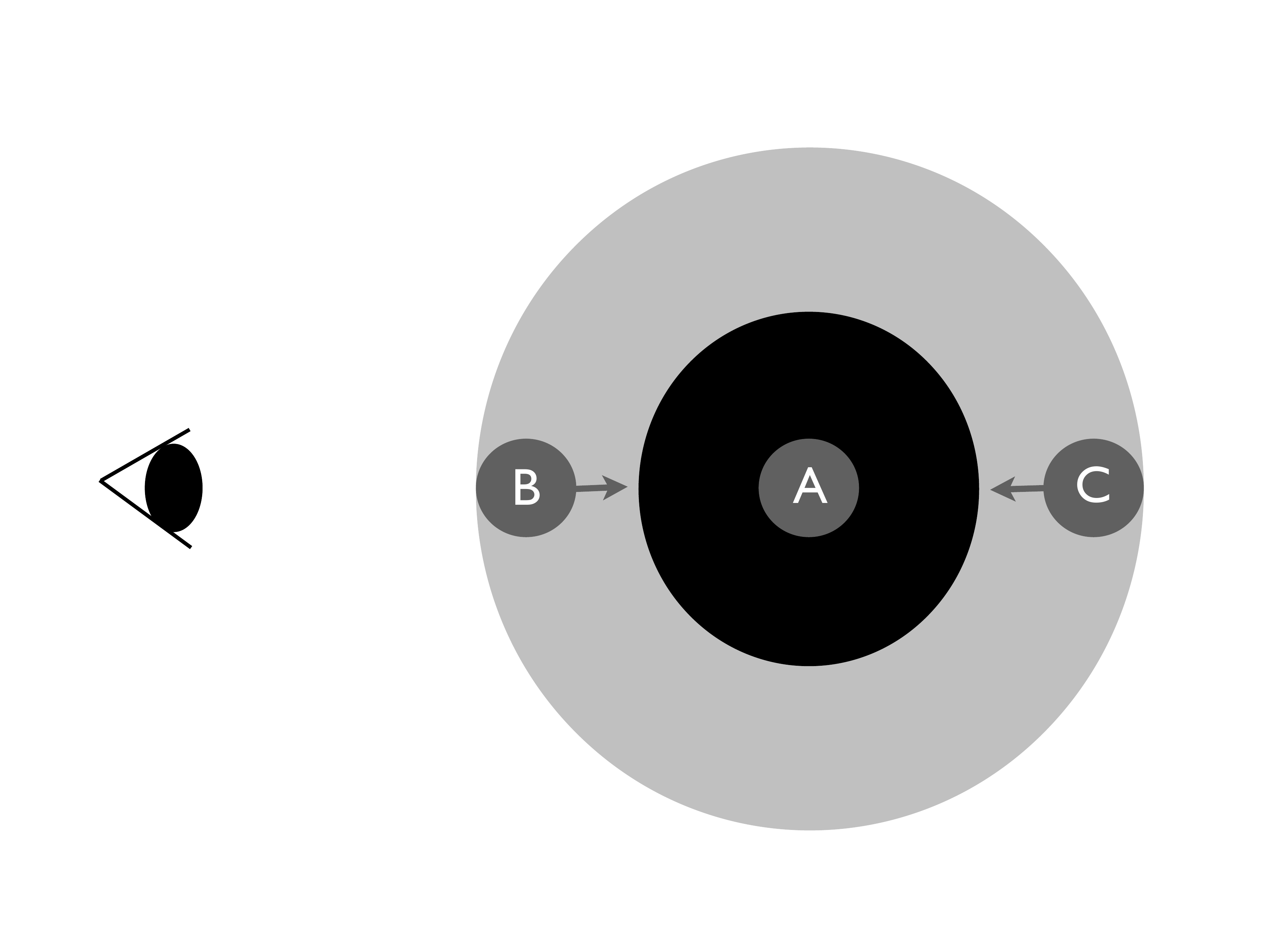}
\vspace{-.5cm}
\caption{Three spherical galaxies of the same physical size and same observed redshift. A is at the centre of a spherical overdensity while B and C are falling towards the centre.}
\end{figure}

The effect is a consequence of the distortion introduced by mapping from redshift-space to real space, as illustrated in Figure 1. Imagine we have three spherical galaxies with the same physical size, and (as an extreme case) the same measured redshift $z_{s}$. Galaxy A is at the centre of a spherical overdensity, and we ignore the contribution from gravitational lensing. A's redshift is purely cosmological, and its angular size is typical for objects at this redshift. Galaxy B is physically nearer to us, with a smaller cosmological redshift, but has a peculiar velocity away from us so that its net redshift is $z_{s}$. Its angular size is therefore larger than typical at this measured redshift. Finally, galaxy C has a larger cosmological redshift and is moving towards us, again with a net observed redshift $z_{s}$. Its angular size is therefore smaller than typical, as it is truly further away. 

The contribution from Doppler lensing to (de)magnification can be summed up as follows:  galaxies with peculiar velocity away from us appear magnified at a particular redshift, relative to typical objects at the same redshift. For galaxies behind an overdensity which are falling towards it, the effect has the opposite sign to gravitational lensing magnification, and is typically much larger than gravitational lensing in the infall region \citep{2013PhRvL.110b1302B}. Similarly, objects behind a void appear magnified~-- opposite to the gravitational lensing contribution.

The Doppler lensing signal is a direct means of measuring velocities in the Universe, and therefore  provides information about the growth rate of structure, a key quantity for discerning between dark energy models and for tests of gravity. In order to exploit the potential of this new probe, we need to find appropriate statistics to measure the effect, and examine the expected signal-to-noise for forthcoming surveys. 

Doppler lensing causes a slight apparent change in size and magnitude for objects at a given observed redshift (throughout we will use the term `size' to mean angular size). However, since these objects have an intrinsic range of sizes and magnitudes, to measure the effect it is necessary to measure size/magnitude for many objects in order to overcome this intrinsic noise. In addition, it is highly desirable to obtain spectroscopic redshifts for the sources, as Doppler lensing between two objects is present over relatively short separations in redshift ($\Delta z \simeq 0.02$), in contrast to gravitational lensing -- which is integrated along the entire line of sight.

Here we suppose that well-calibrated estimates of size and magnitude are available for a catalogue of galaxy images in a survey, which can then be used to obtain noisy estimators of the magnification for each object. Most weak lensing studies so far have used galaxy ellipticities rather than sizes for probing the lensing field. However, size-magnitude estimators have been demonstrated as feasible, and the signal-to-noise  for magnification measurements with these estimators is about half that of shear \citep{2012ApJ...744L..22S}. Once such estimates of magnification have been obtained for a survey, it will be possible to apply the statistics and techniques we develop in this paper in order to measure and use the Doppler lensing effect.

The paper is organised as follows. In Section 2 we review the relevant theory for Doppler lensing, showing the redshift and wavenumber range over which the effect dominates over gravitational lensing. Section 3 describes the simulated data sets that we use to confirm our ability to make measurements of Doppler lensing. In Section 4 we introduce the survey configurations which we consider for our predictions. We then proceed to describe several applications of Doppler lensing: in Section 5, we examine the prospects for detecting the signal around stacked over/under-densities. In Section 6, we calculate suitable power spectra and correlation functions for Doppler lensing, showing that cross-correlation statistics can be measured with high signal-to-noise in future surveys. We consider the impact of intrinsic size correlations and gravitational lensing on these statistics. Section 7 shows how to use Doppler lensing measurements to make 3D maps of a particular potential, related to the gravitational potential but including geometrical factors. We present our conclusions in Section 8.

\section{Doppler lensing: theory}

\subsection{The perturbed $\mathbf \Lambda$CDM model}

The perturbed metric for a $\Lambda$CDM model is 
\be
{ d} s^2  = a^2(\eta)\left[ -(1+2\Phi) {\rm d}\eta^2 +(1-2\Phi){\rm d}\bm{x}^2\right],
\ee
where $\Phi(\eta,\bm{x})$ is the gravitational potential, $a$ is the scale factor, $\eta$ is conformal time and ${x}^i$ are comoving coordinates. At late times, the potential may be written as $\Phi=g(\eta)\Phi_0(\bm x)$, where the growth suppression factor $g(\eta)$ is determined from $g''+ 3 a{H} g' +  a^2\Lambda g= 0$ ($H$ is the Hubble parameter, and a prime is a conformal time derivative: $'=\partial_\eta$) with initial conditions at the end of the radiation era then giving $\Phi_0(\bm x)$, such that $g_0=1$. (To compute the power spectrum of $\Phi$ we use the linear transfer function given in \citet{1998ApJ...496..605E} and to capture non-linear features we use HALOFit~\citep{2003MNRAS.341.1311S}.)

Once $\Phi$ is known, we can find the matter density contrast, $\delta$, and the peculiar velocity $v_i$. The general relativistic Poisson equation gives $\delta$:
\be
\delta =\frac{2a} {3 H_0^2 \Omega_{m} } \left[ \nabla^2\Phi - 3aH\big( \Phi'+aH\Phi \big)\right]\,, \label{grp} 
\ee
where $\Omega_{m}$ is the present day matter density parameter. On small scales the second term in square brackets may be neglected, giving the usual Newtonian expression. The peculiar velocity (of total matter and of galaxies, assuming no velocity bias) is related to $\Phi$ by
\be
v_i = -\frac{2a}{ 3H_0^2\Omega_{m}}\partial_i\big(\Phi'+aH\Phi \big). \label{oi}
\ee

\subsection{Convergence}

The Jacobian relating lensed image positions to unlensed positions is given by \citep[e.g.][]{2001PhR...340..291B}
\begin{equation}
A = \left( \begin{array}{cc} 1-\kappa-\gamma_1&-\gamma_2\\-\gamma_2&1-\kappa+\gamma_1 \end{array}\right)
\end{equation}
where $\kappa$ is the lensing convergence, and $\gamma$ is the lensing shear.  The convergence $\kappa$ includes both gravitational and Doppler lensing, as well as further terms (see equation \eqref{fgr} below); it causes an expansion or reduction of apparent size of an object. The shear $\gamma$ arises principally from gravitational lensing; it causes a change in ellipticity of an object. The distorted apparent angular size of an object $r_l$ is related to its undistorted angular size $ r_u$ by
\begin{equation}
r_l \simeq (1+\kappa)  r_u,
\end{equation}
while the lensed apparent magnitude of an object $m_l$ is related to the unlensed apparent magnitude $ m_u$ by
\begin{equation}
m_l \simeq  m_u + 5 \log_{10}{(1-\kappa)}.
\end{equation}
A simple estimator for the convergence can be derived from an object's measured angular size $r$, which could be derived from fitted parameters  \citep[e.g. the square root of area, ][]{2013MNRAS.433L...6H} or measured using a quadrupole-moment method \citep[e.g.][]{1995ApJ...449..460K}. We can then obtain the mean log size $\langle \ln{r}\rangle$ at redshift $z$, after which a suitable estimator for convergence on a galaxy at redshift $z$ will be
\begin{equation}
\hat{\kappa} = \ln{r}-\langle\ln{r}\rangle_z.
\end{equation}
A more sophisticated estimator, combining size and magnitude, is provided by \citet{2012ApJ...744L..22S}. This estimator is able to take into account the lensing bias, where magnification can bring new faint, small galaxies into the sample. Since galaxies intrinsically have a range of size and brightness, they find (their Figure 1) that the estimator has an intrinsic noise  $\sigma_\kappa\simeq 0.3$, which is the value we adopt throughout this paper.

Now that we have introduced the lensing quantities, we will outline the derivation of the Doppler lensing contribution. 
The lensing convergence $\kappa$ corrects the background angular diameter distance $\bar d_A$ for a source:
\be
d_A(z_s,\bm{n})=\bar d_A(z_s)\big[1-\kappa(z_s,\bm{n})\big]. \label{add}
\ee
The source is \emph{observed} at redshift $z_s$ in the perturbed model and in the direction $-\bm n$, i.e. $\bm n$ is the unit direction vector pointing from the source to the observer. The background area distance at any $z$ is $\bar d_A(z)=\chi(z)/(1+z)$, where $\chi$ is the background comoving distance. The convergence
$\kappa$ may be found by solving the Sachs focusing equation, which follows from the geodesic deviation equation \citep{2001PhR...340..291B}:
\be\label{CovAreadistance}
\frac{d^2 }{d\chi^2}d_A=-{1\over2}  R_{\mu\nu} k^\mu k^\nu\, {d}_A
\ee
(we neglect a second-order shear contribution). Here $\chi$ is the background comoving distance (used as affine parameter along the lightrays), the photon 4-momentum is $k^\mu=dx^\mu/d\chi$ and $R_{\mu\nu}$ is the Ricci tensor. 
We describe a perturbative correction to the angular diameter distance, relative to the background distance $\bar d_A$ at background redshift $\bar z$, by: 
\be\label{d-z-bar}
d_A(\bar z)=\bar d_A(\bar z)\left[1+\frac{\delta d_A}{\bar d_A}\right]\,.
\ee
Solving equation \eqref{CovAreadistance}, we find:
 \begin{eqnarray}
\left. {\delta d_A\over \bar{d}_A }\right|_{\chi_s} &=& -{\bm v}_o\cdot{\bm n} -\Phi_s\nonumber\\ &&\, + \frac{1}{\chi_s} \int_0^{\chi_s} d\chi \big[2\Phi   
  +(\chi-\chi_s) \chi\nabla^2_{\bot}\Phi \big], \label{eq:deltad}
 \end{eqnarray}
where $s$ denotes source and $o$ denotes observer, and ${\bm v}_o$ is the peculiar velocity of the observer. 
The transverse Laplacian (in the screen space orthogonal to the light ray) is given by\footnote{There is a typo in the sign of the last term of equation \eqref{spl} in~\cite{2013PhRvL.110b1302B}. The last term is neglected in \cite{2008PhRvD..78l3530B}, so that $\nabla_\perp^2$ as defined there is not the transverse Laplacian. This does not affect the final result, but it accounts for the difference in appearance between our expression and equation~(31) of \cite{2008PhRvD..78l3530B}.}
\be\label{spl}
\nabla_\perp^2=\nabla^2-(\bm n \cdot \bm\nabla)^2+2\chi^{-1}\bm n \cdot \bm\nabla\,.
\ee
Equation \eqref{eq:deltad} contains Sachs-Wolfe and integrated SW contributions, together with the usual gravitational lensing term
\ba
\kappa_g &=&  \int_0^{\chi_s} { d} \chi
{ (\chi_s - \chi)}{\chi\over\chi_s}  \nabla_\perp^2 \Phi \nonumber\\
&\approx & {3\over2}H_0^2\Omega_m \int_0^{\chi_s} { d} \chi
 (\chi_s - \chi){\chi\over \chi_s}[1+z(\chi)]\delta, \label{kphi}
\ea
where the second line of equation \eqref{kphi} follows on the sub-Hubble scales of interest (see Appendix~A). Equation \eqref{eq:deltad} also includes a Doppler term from the observer's peculiar motion, but no Doppler term associated with the source. 

The Doppler source term comes from the redshift perturbation; the redshift is distorted along a lightray by the volume expansion $\Theta$ and shear $\sigma_{\mu\nu}$ of the matter:
\be
\frac{dz}{d\chi}=-\left[\frac{1}{3} \Theta 
+ \sigma_{\mu\nu}n^{\mu}n^{\nu}\right](1+ z)^2.\label{Covz}
\ee
To linear order, this leads to
\be\label{eq:deltaz}
\left.{\delta z\over 1+\bar z}\right|_{\chi_s}= \big({\bm v}_o-{\bm v}_s\big)\cdot{\bm n}+
\Phi_o- \Phi_s-2 \int_0^{\chi_s} d\chi\,\Phi',
\ee
which contains SW and ISW terms, as well as the Doppler correction from the source's peculiar velocity ${\bm v}_s$. This term is the origin of the Doppler lensing contribution. 

In equations \eqref{eq:deltad} and \eqref{eq:deltaz}, $\chi_s=\chi(z_s)$ is  the co-moving distance calculated in the background spacetime to the source which we infer from the {\em observed} redshift $z_s$, as opposed to the distance to the background redshift $\bar z$ which appears in~\eqref{d-z-bar}. The difference between $\bar z$ and $z_s$ does not affect first-order terms such as \eqref{eq:deltad} and\eqref{kphi} directly,
 as the relevant corrections are second-order (although even for mildly non-linear structures this difference is observable~-- compare the lower panel of Figure \ref{fig:mockcat1} with the upper panel of Figure \ref{fig:mockcat2} below). However, writing \eqref{d-z-bar} in terms of $z_s$ instead of $\bar z$ brings in important extra terms. 
The full perturbation to the angular diameter distance can be written in terms of the {observed} redshift of the source $z_s$ by perturbatively expanding $\bar z$ in \eqref{d-z-bar} and writing $a(\chi_s)=1/(1+z_s)$ so that $H(z_s)={da}/{d\chi}\big|_{\chi_s}$. We then find:
\ba
d_A(z_s)&=&\bar d_A(z_s)\bigg\{1+\left. {\delta d_A\over \bar{d}_A }\right|_{z_s}\nonumber\\
&&+\left[1-\frac{1+z_s}{ H(z_s)\chi_s(z_s)}\right]\left.{\delta z\over 1+\bar z}\right|_{z_s}\bigg\}.
\ea
Then the convergence for a source at observed redshift $z_s$ follows from equation \eqref{add}:
\be
\kappa= \kappa_g +\kappa_v + \kappa_{\rm sw} + \kappa_{\rm isw},
\label{fgr}
\ee
where
\ba
\kappa_v &=& \frac{1+z_s}{H\chi_s}\bm v_o\cdot\bm n+\left(1-\frac{1+z_s}{ H\chi_s}\right)\bm v_s\cdot\bm n\,\label{dlf},\\
\kappa_{\rm sw}&=&2\Phi_s-\Phi_o+\frac{1+z_s}{{H}\chi_s}(\Phi_o-\Phi_s), \label{ksw} \\
\kappa_{\rm isw}&=&-\frac{2}{\chi_s}\int_0^{\chi_s} d\chi\,\Phi+2\left(1-\frac{1+z_s}{ H\chi_s}\right)\int_0^{\chi_s} d\chi\,\Phi',\label{kisw}
\ea
and the gravitational lensing term $\kappa_g$ is given by equation \eqref{kphi}. The SW and ISW terms are generally sub-dominant to the other two contributions and can be neglected, so that
\be
\kappa=\kappa_g +\kappa_v .\label{keff}
\ee
In the Doppler lensing convergence \eqref{dlf},
the term proportional to $\bm v_o\cdot\bm n$ leads to an overall dipole in the magnification. 
We assume that this dipole is subtracted, so that 
\be\label{eq:kappav}
\kappa_v=\left(1-\frac{1+z_s}{ H\chi_s}\right)\bm v_s\cdot\bm n.
\ee
Notice that $\kappa_v$ changes sign depending on whether objects are moving away from or towards us.  Since $\bm n$ is the direction of photon propagation, $\bm v_s\cdot\bm n>0$ for objects moving towards us and $<0$ for objects moving away. At moderate redshift the term in brackets is negative; consequently, 
\begin{itemize}
\item
$\kappa_v<0$ for objects moving towards us -- implying that they appear smaller and dimmer than typical objects at their observed redshift. Their angular distance is higher than inferred from the observed redshift.
\item
$\kappa_v>0$ for objects moving away from us -- they appear larger and brighter than typical objects at their observed redshift. Their angular distance is less than inferred from the observed redshift. In the case of objects at the far end of a void, this magnification is opposite to the demagnification from $\kappa_g$, leading to a significant anti-lensing effect~\citep{2013PhRvL.110b1302B}.
\end{itemize}

With increasing redshift, the factor in brackets in equation (\ref{eq:kappav}) decreases in amplitude, so the magnitude of the Doppler lensing falls -- while that of $\kappa_g$ grows. The factor in brackets goes through zero at the maximum of $\bar{d}_A$, i.e. at $z\sim1.5$ in $\Lambda$CDM. Therefore $\kappa_v$ profiles change sign at high redshift; this is due to an effect which dominates at high redshift, in which the object's image experiences significantly more (or less) cosmic expansion than we inferred from its observed redshift. 

\begin{figure}
\centering
\includegraphics[width=8cm]{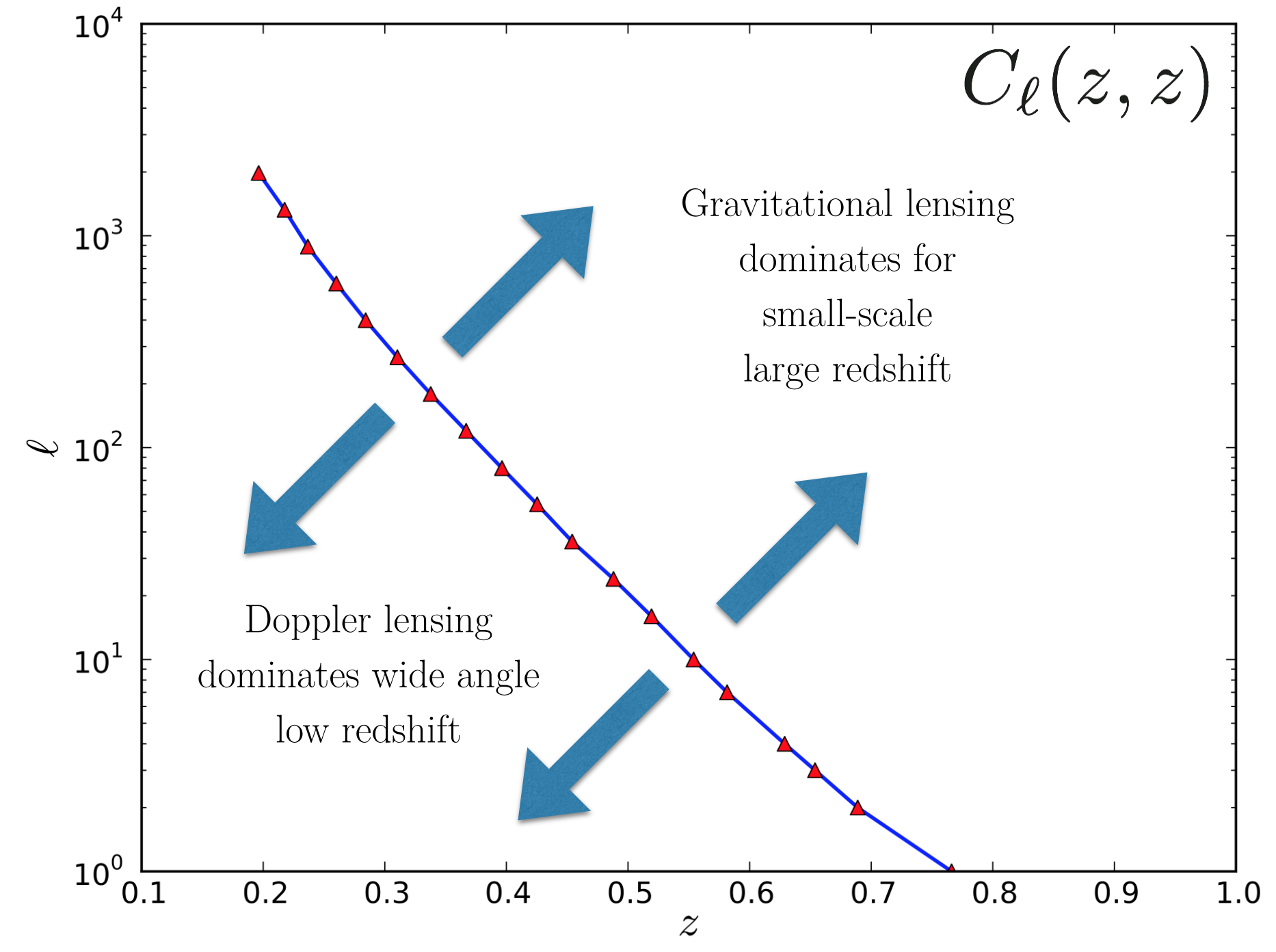}
\caption{The curve shows where the power spectrum for an infinitesimal redshift slice of Doppler lensing ($\kappa_v$) equals that of standard gravitational lensing ($\kappa_g$). Doppler lensing dominates below the curve -- on large scales (small $\ell$) and small redshifts. 
}
\label{fig:scales}
\end{figure}

On what scales is the Doppler lensing important? Using the estimate $|\bm v|\sim  H_0\delta/k$, we expect the effect to be important on large scales. 
The factor in equation \eqref{eq:kappav} is $\mathcal{O}(1)$ for $z\lesssim1$, but its magnitude falls at high redshift. 
The region where the Doppler lensing dominates over standard gravitational lensing is shown in Figure~\ref{fig:scales}. This has been calculated as the points in wavenumber $\ell$ and redshift $z$ where a Doppler lensing power spectrum equals that of a gravitational lensing power spectrum; see Section 6 for details of how we calculate the power spectrum (equation \ref{eq:dopplerpower}). We see that Doppler lensing dominates over gravitational lensing at medium-to-low redshifts and wavenumbers ($\ell \la 1000$ at $z=0.2$, and $\ell \la 100$ at $z=0.4$). As we will see in Section 6.3, the distinct redshift behaviour of Doppler lensing allows us to measure it at much higher redshifts as well. 

\section{Simulations}

For the purpose of testing our Doppler lensing measurement techniques on observational data
we construct two mock galaxy catalogues:
1) a wide-angle $50^{\circ} \times 50^{\circ}$ survey with galaxies having redshifts up to $z =0.3$;
and 2) a deep $10^{\circ}\times 10^{\circ}$ survey up to $z=1$.

These mock catalogues were constructed using the data from 
the Millennium simulation \citep{2005Natur.435..629S,2009MNRAS.398.1150B}.
The Millennium simulation is an N-body simulation for a concordance cosmology
($\Omega_m = 0.25$, $\Omega_\Lambda = 0.75$, $H_0 = 73$ km s$^{-1}$ Mpc $^{-1}$).
It consists of approximately $10^9$ particles of mass $8.6 \times 10^8 M_\odot h^{-1}$
within a cube of volume $(500 h^{-1}$ Mpc$)^3$.
The Millennium simulation is a pure dark matter simulation, but can be populated
with galaxies using semi-analytic galaxy formation models. 
For our mock catalogues we use the semi-analytic galaxies of \cite{2010MNRAS.404.1111G}.
Both data sets are accessible online\footnote{\texttt{
http://gavo.mpa-garching.mpg.de/MyMillennium}}.

The procedure for creating the mock catalogues is as follows:
the observer is placed at the origin (X=0=Y=Z) of the Millennium box.
Because of periodic boundary conditions, if the light ray exits the Millennium box, it enters the other side of the box with entry angles the same as the exit angles.
The boundaries of the light cone are as follows:
the angle between the X and Y axes is set to be between 36 and 86 deg, and
the angle between the Z axis and XY plane is set to be between 2 and 52 deg.
We then use the semi-analytic galaxies data of \cite{2010MNRAS.404.1111G}
(the SQL query \texttt{select snapnum, x,y,z,velX,velY,velZ,r\_mag from Guo2010a..MR}).
If a galaxy lies within the light cone boundaries
(we check for single as well as multiple crossings of the Millennium box, and in addition we compare the time of propagation against the snapshot number) 
then we use the position and velocity data
to calculate the cosmological distance, 
line of sight velocity, and redshifts (both the purely cosmological and
the observed redshift affected by the peculiar velocity).
Then using equation (\ref{eq:kappav}) we calculate the Doppler lensing. To calculate the gravitational lensing we using the dark matter distribution smoothed with a Gaussian kernel of $1.25 h^{-1}$ Mpc
(the SQL query \texttt{select snapnum, phkey, g1\_25 from MField..MField}).
The convergence $\kappa_g$ is evaluated using 
equation (\ref{kphi}). 
Finally, we calculate the observed magnitude
\begin{equation}
 m = m_{\rm rest} + 5 \log_{10} d_L + 25,
\end{equation}
where 
\begin{equation}
d_L(z_s) = (1+z_s)^2 \bar{d}_A(z_s)(1-\kappa_g - \kappa_v).
\end{equation}
For the purpose of our studies we only select galaxies whose observed magnitude is brighter than 26 in the $r$ SDSS band. 

\begin{figure*}
\centering
\includegraphics[width=0.9\textwidth]{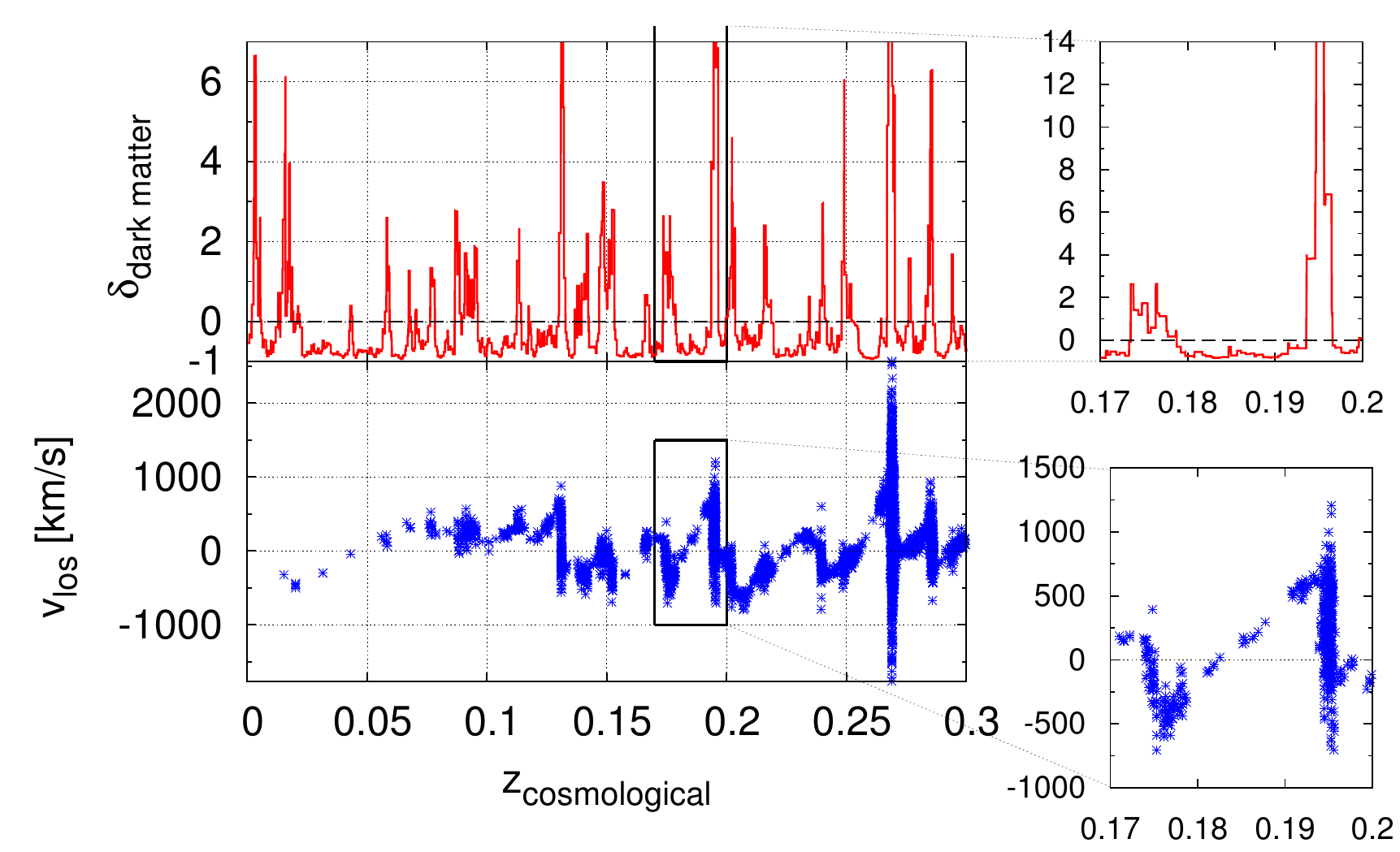}
\caption{
Dark matter distribution ($\delta$) and the line of sight velocity ($v_{\rm los}=-\bm v\cdot\bm n$)
within a narrow light cone of 0.25 sq deg,
as a function of background cosmological redshift $z_\text{cosmological}=\bar z$ (i.e. unaffected by the motion
of galaxies).} 
\label{fig:mockcat1}
\end{figure*}

\begin{figure*}
\centering
\includegraphics[width=0.9\textwidth]{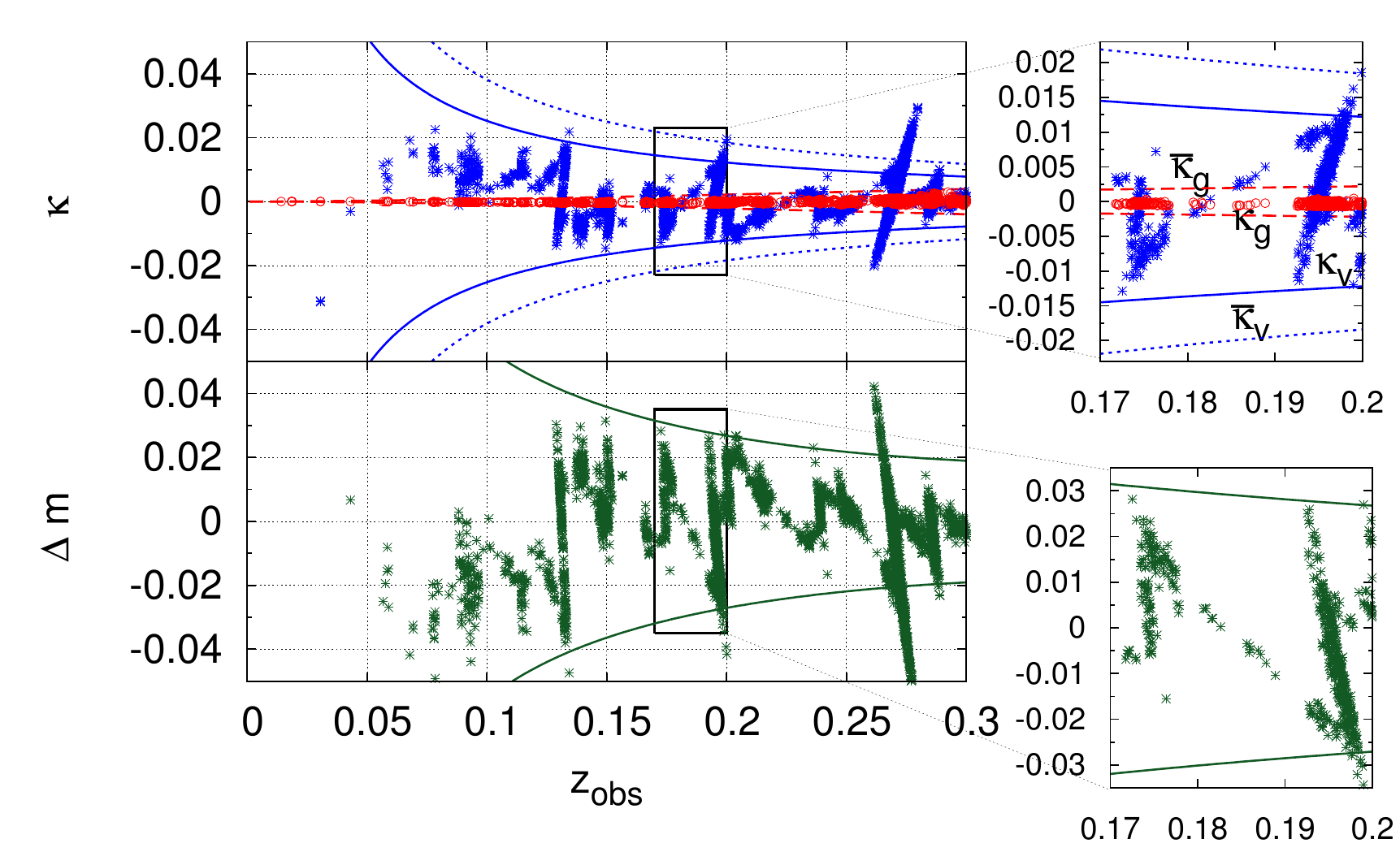}
\caption{Observed convergence ($\kappa$) and the resulting change in magnitude
($\Delta m$),
 within a narrow light cone of 0.25 sq deg,
as a function of the observed redshift 
(i.e. affected by the motion of galaxies). 
The convergence $\kappa$ is presented in the upper panel;
the Doppler convergence $\kappa_v$ is shown using stars,
and the gravitational convergence $\kappa_g$ using open circles.
The dashed line presents the predicted standard deviation of the gravitational lensing signal from equation
(\ref{eq:mkd}), and the solid and dotted lines present the predicted variation of Doppler lensing 
evaluated from equation (\ref{eq:kappav}) with $v_s=355$ km/s and $v_s=535$ km/s respectively.
The change in magnitude $\Delta m = 5 \log_{10} (1- \kappa_v - \kappa_g)$ 
is presented in the lower panel; the solid line show the predicted variation 
$\Delta \bar{m} = 5 \log_{10} [1- (\bar{\kappa}_v^2 + \bar{\kappa}_g^2)^{1/2}]$.
}
\label{fig:mockcat2}
\end{figure*}

An example of a narrow light cone of 0.25 sq deg
extracted from our mock catalogue is presented in Figures \ref{fig:mockcat1} and
\ref{fig:mockcat2}.
Figure \ref{fig:mockcat1} shows the dark matter distribution and line of sight velocities
of galaxies within this narrow light cone.
We can see large variations in both distributions.
The standard deviation of the line of sight velocities of all galaxies in the mock catalogue in the redshift range of 0 to 0.3 is 
approximately $v_{\rm rms} = 355$ km/s, while the average velocity 
(in any direction) is approximately $\bar{v} = 535$ km/s.
As seen from the lower panel, most velocities of galaxies are within this range,
with occasional spikes that are seen around large overdensities in the vicinity of large voids - for example at $z \approx 0.195$ 
and  $ z \approx 0.268$
where the density contrast is $\delta \approx 14.3$ and $\delta \approx 21.5$, respectively,
and the line of sight velocity is 
$v_{\rm los} \approx 1200$ km/s and $v_{\rm los} \approx 2500$ km/s respectively.

As explained in Section 2, both the matter distribution along the line of sight and the peculiar velocities
of galaxies contribute to convergence, and via convergence they affect the observed size and
magnitude of galaxies. These effects are presented in Figure  \ref{fig:mockcat2}.
The gravitational lensing is an integrated effect, and so the induced gravitational convergence is a slowly varying function, without such a large variation as the Doppler convergence, which is a local phenomenon.
We see that for redshifts below 0.3 the Doppler lensing dominates in the convergence,
$\kappa \approx \kappa_v \gg \kappa_\delta$.
Therefore, in this redshift range the convergence $\kappa$
traces the local velocity field rather than the 
integrated density field along the line of sight,
as is expected given the predictions presented in Figure \ref{fig:scales}.
We see that the Doppler lensing also significantly affects the change in magnitude
of galaxies, which is presented 
in the lower panel of Figure \ref{fig:mockcat2} and evaluated 
as 
\begin{equation}
\Delta m = 5 \log_{10} (1- \kappa_v - \kappa_\delta).
\end{equation}
The shape and the amplitude of $\Delta m$ around
cosmic voids seen in Figure \ref{fig:mockcat2} is comparable with that reported by
\cite{2013PhRvL.110b1302B}, whose results were based on analytical models of
single structures embedded in the otherwise homogeneous Universe. Note the behaviour of the convergence and magnitude around non-linear structures, where we obtain steep two-valued functions of redshift. This is caused by the conversion from background redshift to observed redshift (and is a contribution to the Doppler lensing effect which is not captured by our first order treatment in Section 2).

We also compare data from this narrow light cone with the 
expected amplitude of the Doppler lensing $\bar{\kappa}_v$
and gravitational lensing $\bar{\kappa}_g$.
The predicted standard deviation of Doppler lensing is evaluated 
 from equation (\ref{eq:kappav}) and is presented in the 
 top panel of  Figure~\ref{fig:mockcat2}
with solid and dotted lines (for velocity $v = v_{\rm rms} = 355$ km/s and $v = \bar{v} = 535$ km/s respectively). 
The predicted variation of gravitational lensing is evaluated from
\begin{equation} \bar{\kappa}^2_g =
\frac{9 }{4} \Omega_m^2 H_0^4 \int_0^{\chi_s} {\rm d} \chi
\left[ (1+z) \frac{ \chi_s - \chi}{\chi_s} \chi \right]^2
\int_0^{\infty} {\rm d} k ~ k \frac{P(k,z)}{2\pi},
\label{eq:mkd}
\end{equation}
and this corresponds to the dashed curve in the upper panel of Figure~\ref{fig:mockcat2}. 
Similarly, the expected variation of magnitude
\begin{equation}
\Delta \bar{m} = 5 \log_{10} [1- (\bar{\kappa}_v^2 + \bar{\kappa}_g^2)^{1/2}],
\end{equation}
accurately predicts the variation in magnitude within the light cone.
Both the convergence and the change in magnitude seen in the figure
appear to be within reasonable observational limits, and
therefore we might expect to measure them with prospective surveys. We will now turn to the issue of making predictions for the Doppler lensing effect with such surveys.

\section{Prospective Surveys}

In Sections 5 to 7 we will make predictions for measuring Doppler lensing signals with a selection of realistic cosmological survey configurations, representative of forthcoming surveys. We will examine the prospects with three imaging surveys of increasing size, each with dense spectroscopic follow-up:

\vspace{.2cm}
\noindent(i) A 5000 square deg imaging survey, such as that being carried out with the Dark Energy Survey (DES)\footnote{http://www.darkenergysurvey.org/}. We suppose that convergence estimators will exist for all galaxies observed in redshift range $0.1<z<0.3$; we posit dense spectroscopic follow-up for $0.1<z<0.3$ from e.g. 2dF and SDSS overlap regions, plus further spectroscopic redshift campaigns. We assume a number density of objects in the spectroscopic sample of 0.7 per sq arcmin.

\vspace{.2cm}
\noindent(ii) A 15000 square deg imaging survey, such as that planned with the Euclid space telescope\footnote{http://www.euclid-ec.org} \citep{2011arXiv1110.3193L}. Again we include convergence estimators for a sub-sample of the
photometrically observed galaxies, for which we assume we have
spectroscopic redshifts, obtained with a dense follow-up survey. This subsample has redshifts $0.1<z<0.3$ with number density 0.7 per sq arcmin. 

\vspace{.2cm}
\noindent(iii) A 30000 square degree imaging survey, such as the Phase I and/or II Square Kilometre Array (SKA, \citet{2009IEEEP..97.1482D}) could achieve. We choose the same galaxy redshift range and density as before, and suppose that HI spectroscopic redshifts will be available for this galaxy catalogue.

\vspace{.2cm}\noindent
In each of these three cases we assume the same convergence estimator intrinsic noise of $\sigma_\kappa=0.3$ throughout the spectroscopic sample. Notice that in this paper, we are restricting ourselves to a survey in the redshift range $0.1<z<0.3$; this is very conservative, as there is a recoverable Doppler lensing signal at higher redshifts too (see Section 6.3). However, at these higher redshifts, one needs to disentangle the gravitational and Doppler lensing signals; while we sketch the way to achieve this in Section 6.3, we defer detailed predictions for this more complicated case to a later paper.

We now turn to assessing the viability of detecting and utilising Doppler lensing with forthcoming surveys.

\section{Measuring the signal around stacked overdensities}

As an introductory example of Doppler lensing signals which can be measured, we consider the Doppler convergence in a spherical region around an over or under-density, e.g. a void or supercluster~-- this idealised case was shown to give significant Doppler lensing in \citet{2013PhRvL.110b1302B}. 

Averaging the convergence $\kappa_v$ over the objects in the spherical region will lead to a value close to zero, since the Doppler convergence in front of the over- or under-density centre will have the opposite sign to that behind the centre~-- see the inset of Figure \ref{fig:mockcat2}, which shows the signal in front of and behind a void centre, and Figure 1 in \citet{2013PhRvL.110b1302B}. 

To avoid this, we can instead average in the sphere the quantity $\kappa_v \cos{\theta}$, where $\theta$ is the angle between the line of sight and the line connecting the overdensity centre to the lensed galaxy. We will consider a spherically symmetric velocity profile $v(r)$, where $r$ is the distance from the centre of the overdensity, which we assume is much smaller than its distance from the observer. Redshift space distortions will squeeze the sphere into an oblate spheroid in redshift space, but the ellipsoidality of this spheroid is less than $10\%$ on 50Mpc scales and we  neglect this distortion in our locus of averaging here. The component of velocity along the line of sight is $v(r) \cos{\theta}$, so $\kappa_v \cos{\theta}$ is proportional to $\cos^2{\theta}$. Averaging over all spherical annuli of volume $2\pi r^2 \sin{\theta} d\theta dr$ out to radius $R$, we obtain
\begin{equation}
\langle \kappa_v \cos{\theta} \rangle \simeq \left. \frac{A}{3} \int_0^R n(r) v(r) r^2 dr \middle/ \int_0^R n(r) r^2 dr \right.
\label{eq:kvct}
\end{equation}
where $n(r)$ is the 3D number density of objects at radius $r$ and $A=(1-1/a\chi H)$, assumed here to be approximately constant over the radius of the overdensity. The error on $\langle \kappa_v \cos{\theta} \rangle$ is given by
\begin{equation}
\langle \sigma^2 \rangle = \left. \frac{\sigma_\kappa^2}{3} \middle/ \int_0^R n(r) r^2 dr \right.
\end{equation}
where $\sigma_\kappa$ is the intrinsic dispersion on the convergence estimator for an object. In the simple case where the number density of objects is considered to be uniform throughout, denoted $\bar{n}$, the signal-to-noise for measuring $\kappa_v \cos{\theta}$ is
\begin{equation}
	\frac{S}{N} = \frac{2\pi^{1/2}A}{3}\frac{\bar{v}R^{3/2}\bar{n}^{1/2}}{\sigma_\kappa}
	\label{eq:sn}
\end{equation}
where $\bar{v}$ is a typical velocity defined by the ratio of integrals in equation (\ref{eq:kvct}).

\begin{figure}
  \centering
\vspace{-1cm}
\includegraphics[width=8cm]{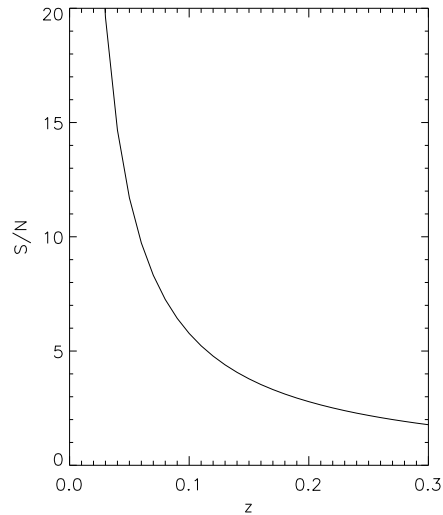}  
\vspace{-2cm}  \caption{Signal-to-noise for measuring the signal for 100 stacked overdensities, with characteristic velocity $\bar{v}=100$kms$^{-1}$ and radius $R=50$Mpc, as a function of redshift.} 
\label{fig:sn}
\end{figure}

We show the resulting signal-to-noise for measuring the $\kappa_v \cos{\theta}$ signal around stacks of 100 over/under-densities in Figure \ref{fig:sn}. Note that here we are examining an optimal situation where these voids are spherical and all of the same size and profile. We show the result for spherical averaging in a radius of 50Mpc, with $\bar{v}=100$kms$^{-1}$; equation (\ref{eq:sn}) shows how to scale for different choices of $\bar{v}$ and radius. We see that it may just be feasible to detect the Doppler lensing in this fashion, with acceptable $S/N$ of $>5$, and it is considerably easier to do this at low redshift (i.e. we do not have to stack so many overdensities). We also see that at no redshift is the measurement feasible for a single void or cluster; stacking would always be necessary. This naturally leads us on to consider other more powerful statistical approaches to measuring Doppler lensing.

\section{Two point statistics}

We now describe suitable two-point statistics for Doppler lensing, which can be measured with forthcoming surveys. We calculate the uncertainties on these statistics for the surveys described in Section 4, and show predictions for cosmological parameters from these statistics. We discuss the systematic effects which could afflict the Doppler lensing measurements, and describe how some of these can be mitigated.

\subsection{Overdensity-convergence cross-correlation}

We will first consider the cross-correlation between the overdensity $\delta$ and the Doppler convergence $\kappa_{v}$. We expect there to be a substantial cross-correlation between these quantities, as overdensities generate the gravitational potential wells which galaxies will fall into with some velocity - and velocity generates Doppler convergence. However, care needs to be taken in constructing a suitable statistic, as velocities from galaxies moving away from us could cancel with velocities moving towards us.

Both the density fluctuation and the Doppler convergence are, at a given redshift, scalar functions on the sky,  so one can expand them in spherical harmonics. In direction on the sky $\bm\theta=-\bm n$ we have
\ba
\kappa_{v}(z,\bm \theta) &=& \sum_{\ell m} \kappa_{\ell m}^{v}(z) Y_{\ell m}(\bm \theta)=-\kappa_{v}(z,\bm n)\,,\\ \nonumber
\delta(z,\bm \theta) &=& \sum_{\ell m}\delta_{\ell m}(z)Y_{\ell m}(\bm \theta)=\delta(z,\bm n)\,.
\ea
The coefficients $\delta_{\ell m}(z)$ and $\kappa_{\ell m}^{v}(z)$ are obtained from
\ba
\delta_{\ell m}(z) &=& \int d\Omega_{\bm \theta}Y_{\ell m}^{*}(\bm \theta)\delta(z,\bm \theta),\nonumber
\\
\kappa_{\ell m}^{v}(z) &=& \int d\Omega_{\bm \theta}Y_{\ell m}^{*}(\bm \theta)\kappa_{v}(z,\bm \theta).
\ea
The angular cross-power spectrum $C^{\delta \kappa_{v}}_{\ell}$ can then be extracted from the average
\be
\label{eq:cdkav}
\langle \delta_{\ell m}(z)\kappa^{v^{*}}_{\ell' m'}(z')\rangle = C^{\delta \kappa_{v}}_{\ell}(z,z')\delta_{\ell\ell'}\delta_{mm'}\,.
\ee
We can also calculate the correlation function between two objects separated by angle $\theta$:
\be
\xi^{\delta\kappa_{v}}(z,z';\theta) = \sum_{\ell = 0}^{\ell_{\rm max}}\frac{2\ell+1}{4\pi}C^{\delta\kappa_{v}}_{\ell}(z,z')P_{\ell}(\theta)\,,\label{corr-fn}
\ee
where $P_{\ell}$ are the Legendre polynomials.
The detailed calculations are presented in Appendix B. We find that the angular cross-power spectrum for a thin shell of galaxy overdensity at redshift $z$ and a thin shell of convergence at redshift $z'$ is given by
\ba
C^{\delta\kappa_{v}}_\ell(z,z')= \frac{16\pi}{\left(3H_{0}^{2}\Omega_{m}\right)^{2}}\int_{0}^{\infty} dk k^2 T^{2}(k)\mathcal{P}_{\Phi_{i}}(k) g(z)b(z)j_{\ell}(k\chi(z))\nonumber\\
\times \left[\left(\frac{1}{\chi(z')  H(z')}-a(z')\right)\mathcal{G}(z')j'_{\ell}(k\chi(z'))\right]
\ea
The definition of $\mathcal{G}(z)$ is provided by equation (\ref{eq:gcal}), $g(z)$ is the growth suppression factor, and $b(z)$ is the linear galaxy bias.

In a real survey, we will examine the cross-correlation in redshift bins of finite width; this is taken into account using window functions $W$ in the radial direction. The cross-power spectrum is then
\ba \label{t1} 
&&\tilde C^{\delta\kappa_{v}}_\ell=\frac{1}{N}\int_{z_{\rm min}}^{z_{\rm max}}dz'W_{1}(z')\int_{z_{\rm min}}^{z_{\rm max}}dzW_{2}(z,z') C^{\delta\kappa_{v}}_\ell( z', z)\nonumber\\
&&= \frac{16\pi}{N\left(3H_{0}^{2}\Omega_{m}\right)^{2}}\int_{0}^{\infty} dk k^{2} T^{2}(k) \mathcal{P}_{\Phi_{i}}(k) \\&&
\times \int_{z_{\rm min}}^{z_{\rm max}}dz'W_{1}(z')g(z')b(z')j_{\ell}(k\chi(z'))\nonumber \\
&&\times  \int_{z_{\rm min}}^{z_{\rm max}}dz W_{2}(z,z') \left(\frac{1}{\chi(z) H(z)}-a(z)\right) \mathcal{G}(z)j'_{\ell}(k\chi(z)) \nonumber
\ea
where $N$ is a normalization factor.  As realistic examples of what could be measured by a survey, we choose (i) a thick bin of width $\Delta z = 0.2$ between $z_{\rm min} = 0.1$ and $z_{\rm max} = 0.3$, and (ii) two thick tomographic bins of width $\Delta z = 0.1$ between $z_{\rm min} = 0.1$ and $z_{\rm max} = 0.3$. The Doppler lensing prevails over gravitational lensing at low and intermediate redshift, and we will therefore neglect the latter. We select the galaxy density fluctuations $\delta_g$ in these thick redshift bins using the window function 
\begin{equation}
W_{1}(z) = n(z)\Theta(z-0.1)(1-\Theta(z-0.3))
\label{eq:w1}
\end{equation}
and the equivalent for the tomographic bins; the galaxy redshift distribution in the bins is approximated as $n(z) \propto z^{2}$ where the proportionality constant is chosen to give the required galaxy density, and $\Theta$ is the Heaviside function. 

We need to be careful in selecting appropriate objects with measured convergences to cross-correlate with these galaxy density fluctuations. Since an overdensity only generates infall velocities over a relatively small distance (or redshift range), we only consider count-convergence pairs for which the two points are within a redshift distance of each other $\Delta z = 0.02$. In addition, we need to avoid averaging Doppler convergences in front of and behind an overdensity, as these cancel each other out (c.f. Section 5). This is achieved by using a one-sided top hat function for the second window function,
\begin{equation}
W_{2}(z,z') = n(z)\Theta(z-z')(1-\Theta(z-z'-0.02))
\end{equation} 
where $z'$ is the redshift of the overdensity, in front of the convergence at redshift $z$. The cross-power for convergence in front of overdensities has the same amplitude and opposite sign, so we will average the absolute value of these two signals, reducing the noise on the power by a factor of $\sqrt{2}$. 

\begin{figure}
  \centering
\includegraphics[width=8cm]{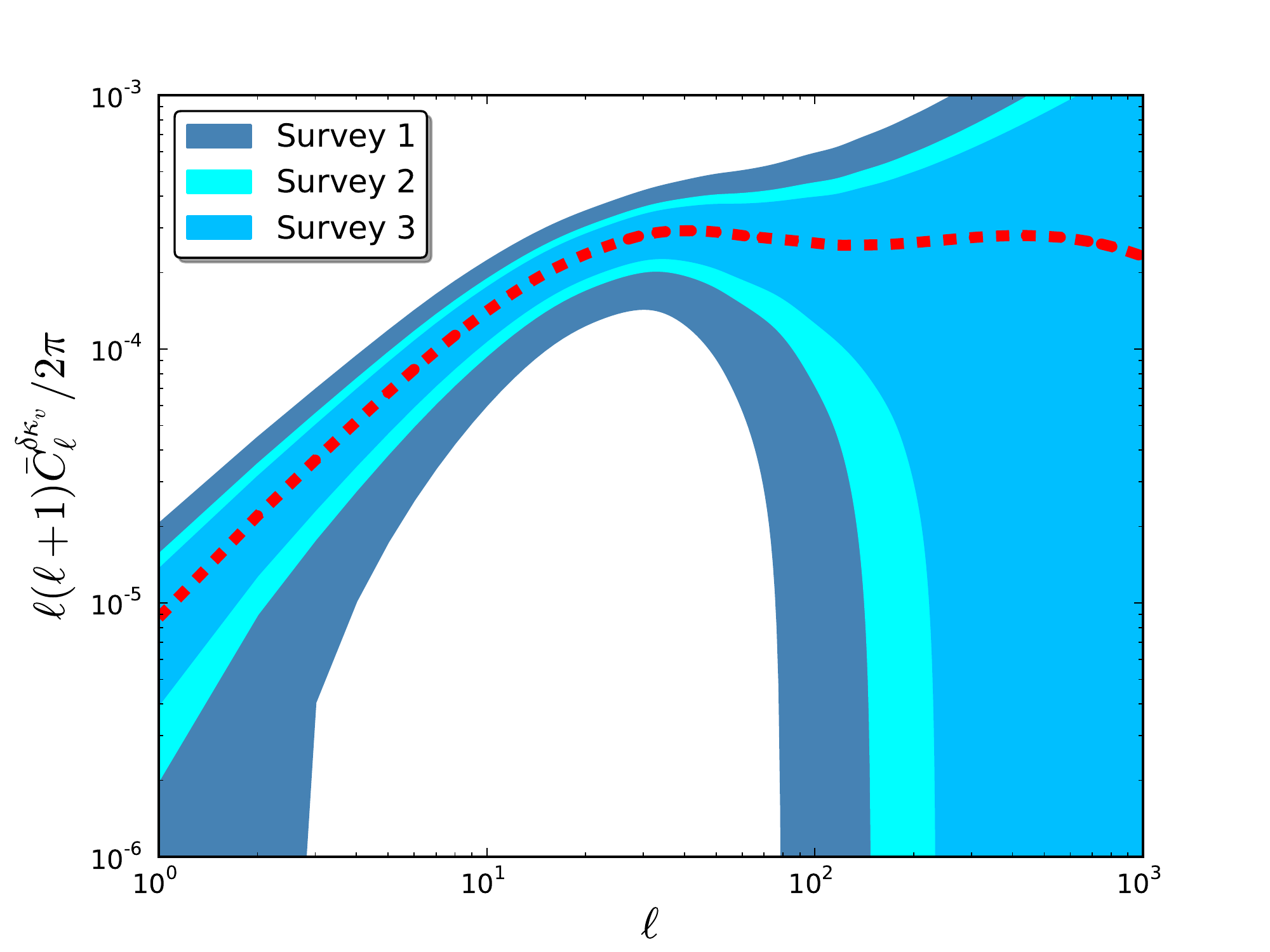} 
\includegraphics[width=8cm]{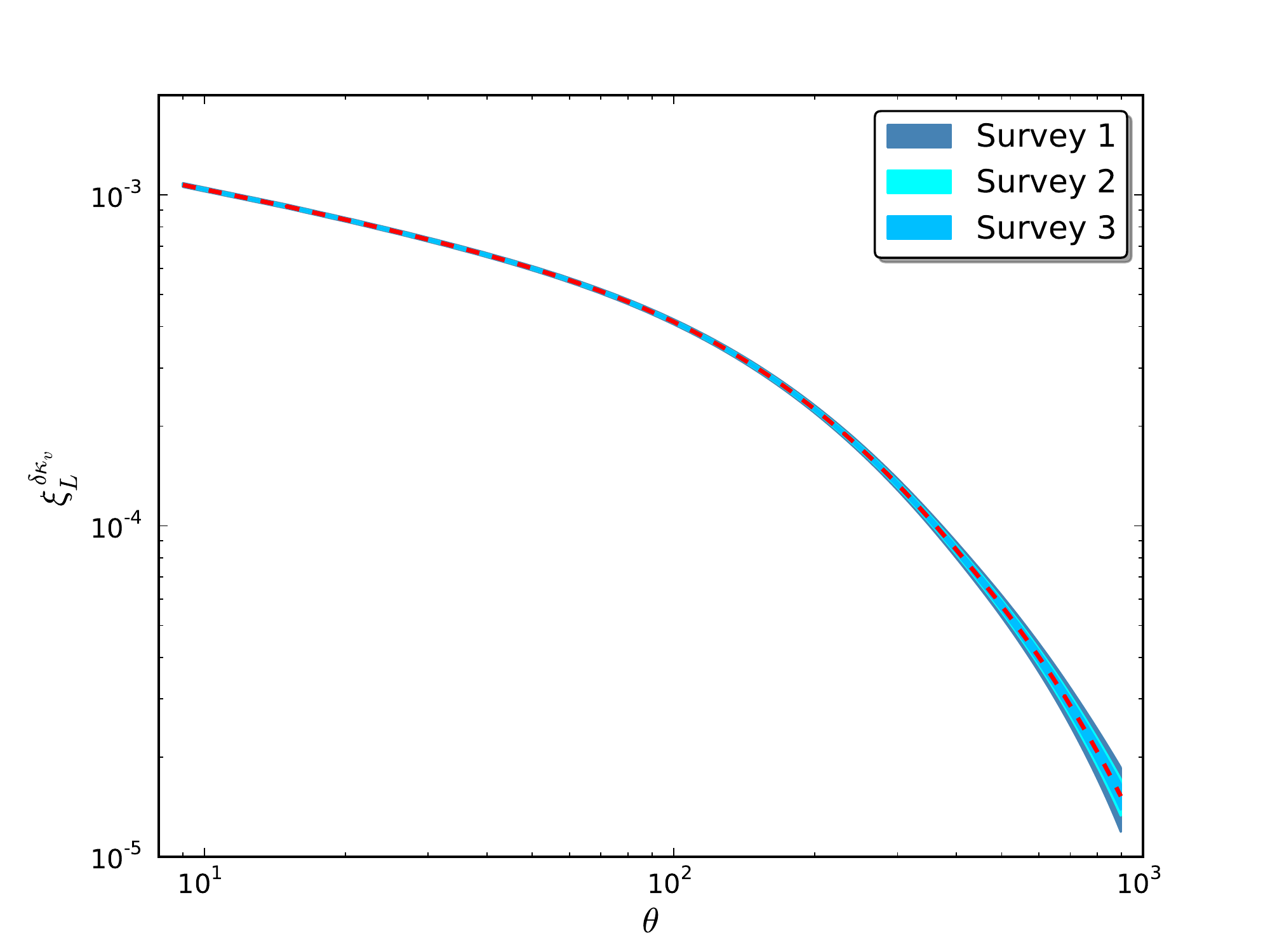} 
  \caption{Top: average angular cross-power spectrum $\bar C^{\delta\kappa_{v}}$.
   The errors are computed by considering the three surveys with sky coverage $f_{\rm sky} = 1/8,3/8,3/4$ 
respectively. Doppler lensing dominates over gravitational lensing for $\ell \la 1000$. Bottom: corresponding average angular cross-correlation $\xi^{\delta\kappa_{v}}$ in real space as a function of angle $\theta$ in arcmin, in angular bins of width $6'$.
 } \label{fig:crosspower}
\end{figure}

As can be noticed from equation \eqref{t1}, the computation is quite expensive as it involves three integrals. 
To circumvent this issue, we subdivide each thick bin into thin shells, so the computation of the cross-correlation for each thin shell involves only two integrals. For each thin shell, we fix the redshift of the overdensities, then compute the two point function between this overdensity slice and the convergences behind it within a bin of width $\Delta z = 0.02$. The angular cross-power spectrum within a thin shell then reads
\ba 
&&C^{\delta\kappa_{v}}_\ell(z')= \frac{1}{N(z')}\int_{z_{\rm min}}^{z_{\rm max}}dz W_{2}(z,z')C^{\delta\kappa_{v}}_\ell(z',z)\\
&&= \frac{16\pi}{N\left(3H_{0}^{2}\Omega_{m}\right)^{2}}\int_{0}^{\infty} dk T^{2}(k)k^{2}g(z')b(z')j_{\ell}(k\chi(z'))\nonumber \\
&&\times  \int_{z_{\rm min}}^{z_{\rm max}}dzW_{2}(z,z') \left(\frac{1}{\chi(z) H(z) }-a(z)\right)\mathcal{G}(z)j'_{\ell}(k\chi(z))\nonumber\\ \label{cw}
\ea
with the normalization factor
\be
N(z') = \int_{z_{\rm min}}^{z_{\rm max}}dzW_{2}(z,z').
\ee
(When we write $C^{\delta\kappa_{v}}_\ell$ with one argument $(z)$ rather than two $(z,z')$ we are referring to equation~\eqref{cw}.)
We finally take the average of all the $C^{\delta \kappa_{v}}_{\ell}(z)$ related to each thin shell to obtain the total angular cross-correlation within a thick bin,
\be
\bar{C}^{\delta\kappa_{v}}_\ell = \frac{\sum z_{i}^{2}C^{\delta\kappa_{v}}_\ell(z_{i})}{\sum z_{i}^{2}}.
\ee
The average angular power spectrum $\bar C^{\delta \kappa_{v}}_{\ell}$ for $0.1<z<0.3$ is shown in Figure \ref{fig:crosspower} (top panel). 
To account for the non-linear evolution of the potential on small scales, the cross-power was computed using the HALOFIT formula \citep{2003MNRAS.341.1311S}, modifying the growth and transfer function. In our calculations, the non-linear matter power spectrum was generated using CAMB \citep{Lewis:1999bs, Howlett:2012mh}, then the effective growth suppression factor becomes dependent on both redshift and wavenumber due to non-linearity, and is computed as 
\be
g_{NL}(z,k) = (1+z)\sqrt{\frac{P_{NL}(z,k)}{P(k)}}
\ee 
where $P_{NL}(z,k)$ is the non-linear matter power spectrum, and $P(k)$ is the present day linear matter power spectrum. At small $k$ (large scales) we have $g_{NL}(z,k) \simeq g(z)$; the non-linear part of the growth suppression factor comes into play when $k$ is large (small scales).

To assess the detectability of the signal, we have computed the error bars for the three different surveys in Section 4, with sky coverage $f_{sky} = 1/8, 3/8,$ and $3/4$ 
respectively. The errors, including cosmic variance and Poisson noise, are estimated as \citep[c.f.][]{2004PhRvD..70d3009H}
\ba
\Delta \bar{C}_{\ell}^{2} &=& \frac{1}{(2\ell+1)f_{sky}}\left[\left(\bar{C}^{\delta\kappa_{v}}_{\ell}\right)^{2}\nonumber\right.\\&&\left.+\left(\bar{C}^{\kappa_{v}\kappa_{v}}_{v}+\frac{\sigma_{\kappa}^{2}}{n_{\kappa}}\right)\left(\bar{C}^{\delta \delta}_{\ell}+\frac{1}{n_{g}}\right)\right]
 \ea
where $n_\kappa$ is the number density of the thin convergence bin and $n_g$ is the number density of the thick bin in question. Figure \ref{fig:crosspower} (top panel) shows the errors (shaded area) corresponding to each survey. The signal can be measured over a wide range in wavenumber in all three survey configurations; note that the error is on each individual $\ell$, so band averaging can be used to measure the signal to high $\ell$. The errors are large on low $\ell$ modes due to cosmic variance, and are large at high $\ell$ due to the limited number of objects at small scales. In addition, we expect that gravitational lensing will dominate over the doppler lensing signal presented here for $\ell \ga 1000$ (see figure \ref{fig:scales}).

We can now calculate the angular correlation function in real space, using equation~\eqref{corr-fn}. 
To compute the errors on this quantity, we use
\begin{equation}
\sigma_{\xi}^{2}(\theta_{i},\Delta\theta) \simeq \sum_{0}^{\ell_{max}} \Delta \bar{C}_{\ell}^{2}\left(\frac{2\ell + 1}{4\pi}\right)^{2}\frac{P_{\ell}(\theta_{i})+P_{\ell}(\theta_{i+1})}{\Delta \theta^{2}}
\label{eq:xierr}
\end{equation}
where $\Delta \theta$ is the angular bin size; here we choose $\Delta \theta = 6'$. The correlation function and errors for different surveys, for one thick redshift slice $0.1<z<0.3$, are shown in the lower panel of Figure \ref{fig:crosspower}; notice the small error bars on this statistic for each of our prospective surveys.

\begin{figure}
\hspace{-.5cm}
\includegraphics[width=10cm]{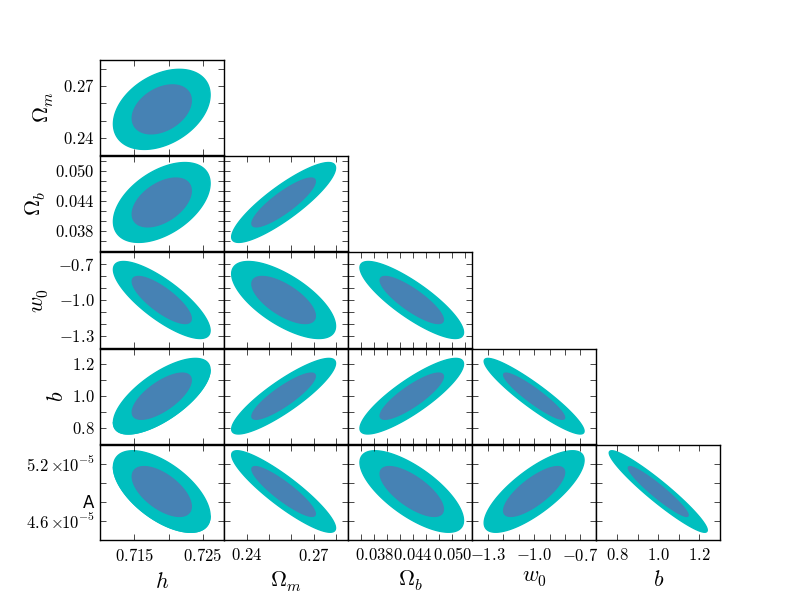}  
 \caption{68\% (blue) and 95\% (light blue) confidence ellipses for cosmological parameters. Here, survey (iii) is used to compute the Fisher matrix, with two low redshift tomographic bins.}
\label{fig:fisher}
\end{figure}

We can obtain from these statistics estimates of constraints on cosmological parameters, using the Fisher information matrix. This is given here by \citep[c.f.][]{2004PhRvD..70d3009H}
\begin{equation}
F_{\alpha \beta} = f_{\rm sky}\sum_{\ell} (2\ell + 1)\frac{\partial \bar{C}^{\delta \kappa_{v}}_{\ell}}{\partial X_{\alpha}}
\frac{\partial \bar{C}^{\delta \kappa_{v}}_{\ell}}{\partial X_{\beta}}\frac{1}{\Delta \bar{C}_{\ell}^{2}}
\end{equation}
where $X_\alpha$ is the set of cosmological parameters we consider. Here we choose ${\bm X} = \{A, b, h, \Omega_b,\Omega_m,w_0 \}$, where $A$ is the primordial power spectrum amplitude, $b$ is the linear galaxy bias, $h$ is the Hubble parameter, $\Omega_b$ is the baryon density parameter, $\Omega_m$ is the total matter density parameter, and $w_0$ is the dark energy equation of state. We calculate this Fisher matrix using wavenumbers $1<\ell<1000$.
Assuming the parameter likelihood is approximately a multivariate Gaussian around its peak, the Fisher matrix is the inverse of the covariance matrix of the parameters; in Figure \ref{fig:fisher}, we show the resulting 68\% and 95\% CL ellipses for pairs of cosmological parameters, where the other cosmological parameters have been marginalised over. We show the results for the two tomographic bin case, for survey (iii) with no other cosmological information (e.g. Planck or supernovae priors). We see that the constraints are very promising, bearing in mind that this is for only two low redshift tomographic bins; we obtain marginalised error on dark energy equation of state of $\sigma_w=0.13$.
We pursue the question of whether we can push our measurements to high redshift in Section 6.3. 

In order to demonstrate the practicality of measuring this signal, we have measured the cross-correlation function $\xi^{\delta \kappa_v}$ in our $50^\circ \times 50^\circ$ simulated dataset (see Section 3). We use the same window functions as above: we include galaxies with redshifts $0.1<z<0.3$, dividing these into two bins with $0.1<z<0.2$ and $0.2<z<0.3$.  We average the galaxy counts and Doppler convergence estimators in pixels with size $1^\circ$ transversely and 0.001 in the redshift direction, calculating $\delta_g$ as the overdensity of counts in pixels. We then calculate the correlation function for the $j$th tomographic bin,
\be
\xi^{\delta \kappa_v}(\theta_i, z_j)= \sum_A \delta(z,\vec{\theta}) \kappa_v(z',\vec{\theta}')-\sum_B \delta(z,\vec{\theta}) \kappa_v(z',\vec{\theta}')
\label{eq:xideltkap}
\ee
where sum $A$ is over $\delta$ pixels in the $j$th thick tomographic bin, and over $\kappa_v$ pixels with $z'>z$, within 0.02 in redshift of the $\delta$ pixel in the pair, and with appropriate angular separation to be in the $\theta_i$ bin. Sum $B$ is over $\delta$ pixels in the thick tomographic bin, and over $\kappa_v$ pixels with $z'<z$, within 0.02 in redshift of the $\delta$ pixel in the pair, and with appropriate angular separation to be in the $\theta_i$ bin.

\begin{figure}
  \vspace{-3.2cm} \hspace{-1cm}
\includegraphics[width=10cm]{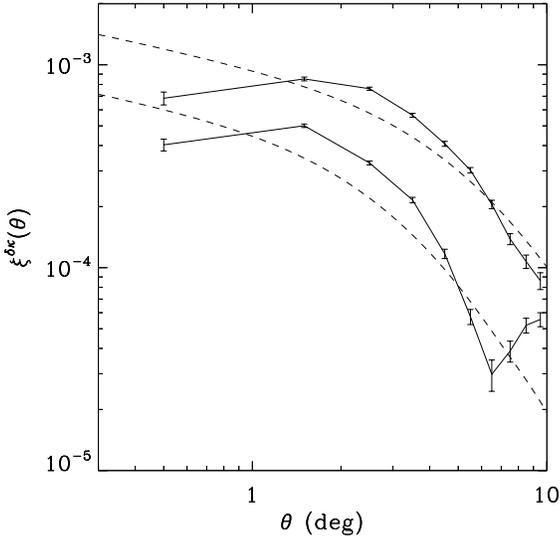}  \vspace{-3cm}
  \caption{Overdensity - doppler convergence cross-correlation function as a function of separation angle. The solid lines show the correlation measurements for the tomographic bins $0.1<z<0.2$ (upper line) and $0.2<z<0.3$ (lower line), and the dashed lines show  the theoretical predictions.} 
\label{fig:deltkapsim}
\end{figure}

We show our resulting correlation function in two tomographic bins in Figure \ref{fig:deltkapsim}. 
We see that the simulated measurements are similar in magnitude to our theoretical calculations, and are precisely measured even when splitting the signal tomographically. Deviations from our theoretical curves will be reduced in future by more detailed modelling of the nonlinear velocity field, and taking into account the slope effect seen in Figure \ref{fig:mockcat2}.

\subsection{Doppler convergence autocorrelation}

Having calculated the cross-correlation between the overdensity $\delta$ and the Doppler convergence $\kappa_{v}$, it is also of interest to calculate the power spectrum and auto-correlation for Doppler convergence. Using the definition 
\be
\label{eq:ckkav}
\langle \kappa^{v}_{\ell m}(z)\kappa^{v^{*}}_{\ell' m'}(z')\rangle = C^{\kappa_{v}\kappa_{v}}_{\ell}(z,z')\delta_{\ell\ell'}\delta_{mm'}
\ee
we find the Doppler convergence power spectrum to be (see Appendix B for derivations)
\ba
&&\bar{C}^{\kappa_{v}\kappa_{v}}_\ell= \frac{16\pi}{N\left(3H_{0}^{2}\Omega_{m}\right)^{2}}\int_{0}^{\infty} dk kT^{2}(k)
\mathcal{P}_{\Phi_{i}}(k)\label{eq:dopplerpower}\\&&
\times \left[\int_{z_{\rm min}}^{z_{\rm max}}dz \left(1-\frac{1}{\chi aH}\right)a(z)W_{1}(z)\mathcal{G}(z)j'_{\ell}(k\chi(z))\right]\nonumber\\&& 
\times \left[\int_{z_{\rm min}}^{z_{\rm max}}dz' \left(1-\frac{1}{\chi aH}\right)a(z')W_{2}(z',z)\mathcal{G}(z')j'_{\ell}(k\chi(z'))\right]\nonumber
\ea
\citet{2008PhRvD..78l3530B} used the continuity equation involving the density fluctuations to compute this Doppler convergence power spectrum,
whereas in our case we have used the potential $\Phi$ and have recovered the same results.
 
To show the feasibility of measuring Doppler lensing correlations, we make calculations for the three surveys in Section 4, choosing the same thick redshift bins as in Section 6.1: (i) a thick bin of width $\Delta z = 0.2$ between $z_{\rm min} = 0.1$ and $z_{\rm max} = 0.3$, and (ii) two thick tomographic bins of width $\Delta z = 0.1$ between $z_{\rm min} = 0.1$ and $z_{\rm max} = 0.3$. We correlate all the objects at a given redshift $z'$ with all the objects behind within a distance of $\Delta z = 0.02$, and we average over the thick bin in question. 
For these configurations, $W_{1}(z)$ has the same form as equation (\ref{eq:w1}) and $W_{2}(z)$ is given by
\begin{equation}
W_{2}(z,z') = n(z)\Theta(z-z')(1-\Theta(z-z'-0.02))
\end{equation} 
Again, since the computations are expensive due to the three integrals, we adopt the same approach we used for the cross-correlation function; the Doppler convergence power within a thin shell is given by
\ba
&&C^{\kappa_{v}\kappa_{v}}_\ell(z')= \frac{16\pi}{N\left(3H_{0}^{2}\Omega_{m}\right)^{2}}\int_{0}^{\infty} dk kT^{2}(k)
\mathcal{P}_{\Phi_{i}}(k)\\&&
\times \left[\left(1-\frac{1}{\chi aH}\right)a(z')\mathcal{G}(z')j'_{\ell}(k\chi(z'))\right]\nonumber \\&& 
\times \left[\int_{z_{\rm min}}^{z_{\rm max}}dz\left(1-\frac{1}{\chi aH}\right)a(z)W_{2}(z',z)\mathcal{G}(z)j'_{\ell}(k\chi(z))\right]\nonumber
\ea 
with the normalization factor
\begin{equation}
N(z') = \int_{z_{\rm min}}^{z_{\rm max}}dzW_{2}(z,z').
\end{equation}
Then the average Doppler convergence power within a thick bin is found using
\begin{figure}
  \centering
\includegraphics[width=8cm]{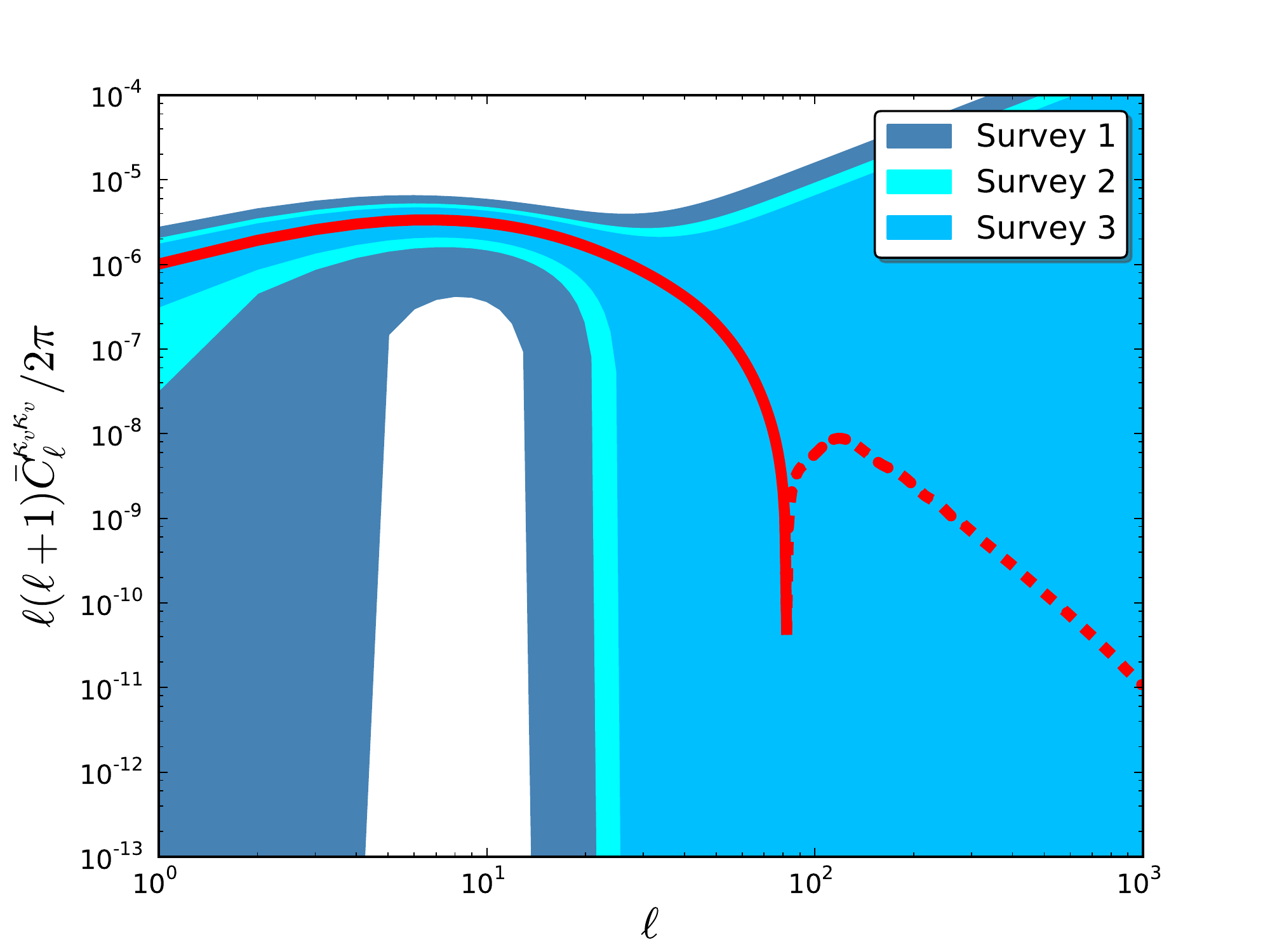}  
\includegraphics[width=8cm]{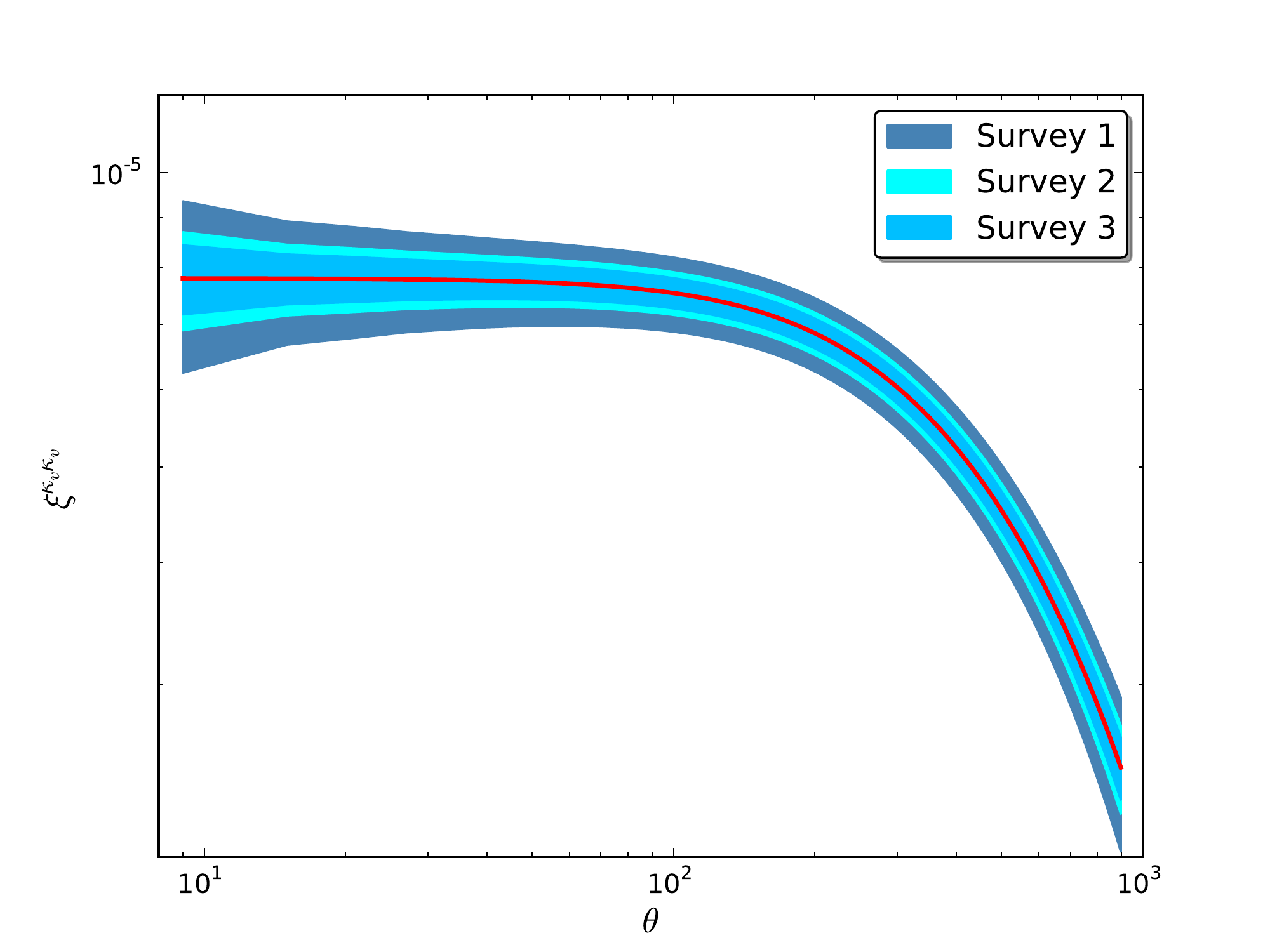}  
  \caption{Top: Power spectrum of the Doppler convergence in harmonic space. The dashed line indicates negative values, whereas the solid line indicates positive values. Bottom: Correlation function of the Doppler convergence in real space.} 
 \label{fig:ckk}
\end{figure}
\be
\bar{C}^{\kappa_{v}\kappa_{v}}_\ell = \frac{\sum z_{i}^{2}C^{\kappa_{v}\kappa_{v}}_\ell(z_{i})}{\sum z_{i}^{2}}.
\ee
Figure \ref{fig:ckk} shows the resulting angular power spectrum $\bar{C}^{\kappa_{v}\kappa_{v}}_\ell$ for redshift averaging (i). At low wavenumber (large physical scales), the signal is positive, indicating that most pairs at large separation in this redshift slice are on the same side of the nearest large-scale gravitational well, which may extend outside of the slice. At high wavenumber (small physical scales), we have negative values indicated by the dashed line; this is due to there being an excess of pairs in our sample on these scales which have velocities facing towards each other (due to there being many small-scale potential wells within the redshift slice for which pairs can exist on either side). We compute the errors in the power spectra for our three surveys; 
the errors in the $\bar{C}_{\ell}$ are given by
\begin{equation}
\Delta \bar{C}_{\ell}^{2} = \frac{1}{(2\ell+1)f_{\rm sky}}\left(\bar{C}^{\kappa_{v}\kappa_{v}}_{\ell}+\frac{\sigma_{\kappa}^{2}}{n_{\rm thin}}\right)\left(\bar{C}^{\kappa_{v}\kappa_{v}}_{\ell}+\frac{\sigma_{\kappa}^{2}}{n_{\rm thick}}\right)
 \end{equation} 
where $n_{\rm thin}$ and $n_{\rm thick}$ are the galaxy number densities of the thin shell and the thick bin respectively. We see in Figure \ref{fig:ckk} that the Doppler convergence power spectrum can be measured on a range of scales; note again that the errors shown are for each individual $\ell$, so averaging in band powers will lead to tighter error bars on each bin.

We have also computed the autocorrelation function $\xi^{\kappa_v \kappa_{v}}(\theta)$ in real space using the relation
\be
\xi^{\kappa_v \kappa_{v}}(\theta) = \sum_{\ell = 0}^{\ell_{\rm max}}\frac{2\ell+1}{4\pi}\bar{C}^{\kappa_v \kappa_{v}}_{\ell}P_{\ell}(\theta)
\ee
The bottom panel of Figure \ref{fig:ckk} shows the Doppler convergence correlation in real space for redshift averaging (i). Again, the error bars are computed for the three surveys defined in Section 4, using equation (\ref{eq:xierr}), choosing a bin size $\Delta\theta = 6'$. We see that the autocorrelation function is measurable with each of the three surveys, but at lower signal-to-noise than the cross-correlation function $\xi^{\delta \kappa_v}$.

\begin{figure}
  \vspace{-3.2cm} \hspace{-1cm}
\includegraphics[width=10cm]{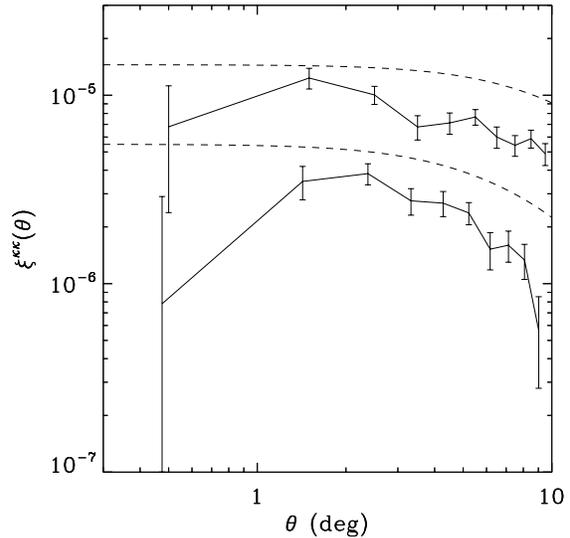}  \vspace{-3cm}
  \caption{Doppler convergence autocorrelation function as a function of separation angle. The solid lines show the correlation measurements for the tomographic bins $0.1<z<0.3$ and $0.2<z<0.3$, and the dashed lines show the theoretical prediction.} 
  \label{fig:kapkapsim}
\end{figure}

We use our $50^\circ \times 50^\circ$ simulated dataset (see Section 3) to attempt to measure the autocorrelation $\xi^{\kappa_v \kappa_v}$. We use the same window functions as above: we include two redshift bins with $0.1<z<0.2$ and $0.2<z<0.3$. Using the same pixelisation as in Section 6.1, we calculate the correlation function for the $j$th tomographic bin,
\be
\xi^{\kappa_v \kappa_v}(\theta_i, z_j)= \sum_A \kappa_v(z,\vec{\theta}) \kappa_v(z',\vec{\theta}')
\ee
where sum $A$ is over $\kappa_v$ pixels in the $j$th thick tomographic bin, and over $\kappa_v$ pixels within 0.02 in redshift (either side) of the first pixel in the pair, and with appropriate angular separation to be in the $\theta_i$ bin. 

We show our resulting correlation function in two tomographic bins in Figure \ref{fig:kapkapsim}.
We are able to measure the signal between 1$^\circ$ and $10^\circ$ on this 2500 sq deg survey, in two low redshift bins. Our measurements are of a similar amplitude to the predicted values, but are somewhat suppressed in both tomographic bins; further modelling of the nonlinear velocities and the slope effect of Figure \ref{fig:mockcat2} are clearly required for more accurate predictions.

\subsection{The impact of intrinsic size/brightness correlations and gravitational lensing}

In the above analysis, we have neglected two potential contributors to our correlation functions: intrinsic size/brightness correlations between galaxies, and the gravitational lensing effect. Here we will discuss how these two effects impact our results. For simplicity, we will only discuss size throughout this section, although all the same arguments pertain for magnitude.

\subsubsection{Intrinsic correlations}

It is well known that gravitational lensing shear studies suffer from a systematic effect, the intrinsic alignment of galaxy ellipticities (see e.g. \citet{2013MNRAS.436..819J}). Two effects contribute: the `II' effect where two background galaxies are physically aligned with each other, mimicking the gravitational shear signal; and the `GI' effect where a background galaxy is lensed by a foreground halo, which is also affecting the orientation of a physically nearby galaxy. 

It is plausible that similar intrinsic correlations exist for the size of objects. The equivalent of the `II' effect is the possibility that galaxies which are near each other and in a similar environment may have correlated physical sizes. There are two effects equivalent to the `GI' effect: the first is the idea that a background object may have gravitational lensing convergence because of a foreground halo, which is also contributing to an environmental dependence on another galaxy's size in its vicinity. The second is the idea that an object may have Doppler convergence due to a nearby halo, which is also contributing to an environmental dependence on another galaxy's size in its vicinity. Will these effects cause problems for Doppler lensing measurements?

We can analyse the issue symbolically in the following way. The excess size of an object can be described by a total estimated convergence,
\begin{equation}
\kappa=\kappa_{\rm int} + \kappa_v + \kappa_g
\end{equation}
where $\kappa_{\rm int}$ is a measure of the fact that a galaxy may be intrinsically larger or smaller than usual. $\kappa_v$ and $\kappa_g$ are the Doppler and gravitational lensing convergence respectively.

If we autocorrelate the total convergence, we obtain
\begin{eqnarray}
\langle \kappa \kappa \rangle&=&\langle \kappa_{\rm int}\kappa_{\rm int} \rangle + 2 \langle \kappa_v \kappa_{\rm int}\rangle+ 2\langle \kappa_g \kappa_{\rm int} \rangle\nonumber\\
 &&+ 2 \langle \kappa_g \kappa_v \rangle +  \langle \kappa_g \kappa_g \rangle + \langle \kappa_v \kappa_v \rangle
 \label{eq:kk}
\end{eqnarray}
The first term here is the II term; the next two terms are the two GI terms; the sixth term is the Doppler auto-correlation that we are interested in. We will discuss the fourth and fifth terms in the next subsection.

Of the first three terms, we can immediately discount the third for the Doppler lensing statistics discussed above. The source of such a correlation would be a halo near a galaxy with $\kappa_{\rm int}$, which also gravitationally lenses the other galaxy in the correlation function pair. Our Doppler lensing auto- and cross-correlations are designed to only include pairs separated by a small distance in redshift ($\Delta z = 0.02$), over which distance the growth of gravitational lensing convergence is insignificant.  

However, the first and second terms need not be small; as in gravitational lensing shear studies, it will be necessary to develop a model for intrinsic size correlations, and fit jointly for Doppler convergence and size correlation parameters when making cosmological constraints with the autocorrelation function.

The situation is much simpler in the case of the overdensity-Doppler cross-correlation. Here, the terms are 
\begin{equation}
\langle  \kappa \delta \rangle=\langle \kappa_{\rm int} \delta \rangle + \langle \kappa_v \delta \rangle+ \langle \kappa_g \delta \rangle
\label{eq:kd}
\end{equation}
The term we are interested in is the second term; we will consider the third term in the next subsection. The first term is the systematic intrinsic size term, which for some choices of correlation function could be large - the intrinsic size of galaxies plausibly depends on the density of their environment. However, we have chosen to average pairs with opposite signs depending on whether  the $\kappa$-galaxy is in front of or behind the $\delta$-galaxy. Since $\kappa_{\rm int}$ has no knowledge of whether it is in front of or behind an overdensity, this term averages to zero. 

Hence our cross-correlation is not affected by intrinsic size correlations. The same is true of the statistic in Section 5, which again averages convergence with opposite signs in front of and behind an overdensity centre (via the $\cos \theta$ factor).

\subsubsection{Gravitational lensing}

In addition to the intrinsic size correlation, a further effect can mix with the Doppler lensing signal: gravitational lensing itself. We see this in equation (\ref{eq:kk}), where the fourth and fifth terms involve gravitational lensing convergence. 

The fourth term describes correlations between Doppler and gravitational lensing. This can be made large for some configurations - for instance if the average is over pairs selected to be where one object is just behind an overdensity (and hence Doppler lensed), and the other is far behind the same overdensity (and hence gravitationally lensed). However, for the average suggested in Section 6.2, the term will be very small due to the thin $\Delta z=0.2$ redshift shell - a halo causing Doppler convergence in a nearby galaxy cannot cause substantial gravitational lensing in such a thin shell.

The fifth term is negligible at low redshifts (c.f. Figure \ref{fig:scales}), and hence does not impact on our calculations in Section 6.2 made for $0<z<0.3$. However, at higher redshifts, this term will dominate over the Doppler autocorrelation. In this case, it is still possible to distinguish the gravitational lensing and Doppler lensing signals, due to the fact that while convergence is caused by Doppler and gravitational lensing, shear is only caused by gravitational lensing to a very good approximation \citep[see][]{2008PhRvD..78l3530B}. In addition, the two point statistics of gravitational shear and gravitational convergence are identical \citep[e.g.][]{2001PhR...340..291B}. Hence, if we have estimators for total convergence $\kappa$ and shear $\gamma$ for a set of objects, and if intrinsic alignments and intrinsic size correlations have been fully modeled and subtracted, we have
\begin{eqnarray}
\langle \kappa \kappa \rangle&=&  \langle \kappa_g \kappa_g \rangle + \langle \kappa_v \kappa_v \rangle \\
\langle \gamma \gamma \rangle &=& \langle \kappa_g \kappa_g \rangle
\end{eqnarray}
Therefore the shear correlation function gives us the pure gravitational lensing signal, and $\langle \kappa \kappa \rangle-\langle \gamma \gamma \rangle$ gives us the pure Doppler lensing signal.

Again the situation is much simpler with the cross-correlation, equation (\ref{eq:kd}). While the third term here can be large for a thick redshift bin with uniform averaging, this is not the configuration chosen in Section 6.1. There, the average is over a thin slice, and has opposite signs depending on whether the $\kappa$ object is in front of or behind the $\delta$ object. Hence we expect this term to average to a small value (the typical lensing due to an overdensity at a redshift separation $<0.02$ behind the overdensity) even at high redshift, leaving only the second term which is the Doppler cross-correlation of interest. 

\begin{figure}
  \vspace{-3cm} \hspace{-.5cm}
\includegraphics[width=10cm]{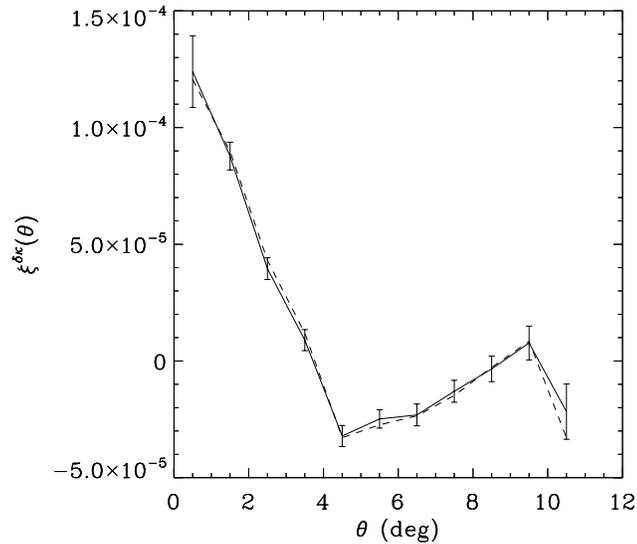}  \vspace{-3cm}
  \caption{Overdensity - doppler convergence cross-correlation function as a function of separation angle, for a thick $\delta$ bin at $0.5<z<0.9$. Solid line shows the cross-correlation when gravitational lensing has not been added to the simulation; dashed line shows the cross-correlation including gravitational lensing.} \label{fig:highdelta}
\end{figure}

We have checked that this argument is correct using the deep $10^\circ \times 10^\circ$ simulation described in Section 3. We have measured the cross-correlation function given in equation (\ref{eq:xideltkap}), in a thick $\delta$ bin of $0.5<z<0.9$, with a thin $\kappa$ region of $\Delta z = 0.02$, and opposite signs on the average when $\kappa$ is in front of or behind $\delta$. We measure the cross-correlation function for our simulation with gravitational lensing switched off, then with gravitational lensing present. The results for these two cases are shown in Figure \ref{fig:highdelta}; we see that gravitational lensing does not appear to have a significant effect on this statistic, even at high redshift. Nevertheless, for future precision cosmological measurements, a joint model of Doppler and gravitational lensing will remove any remaining bias.

\section{Maps}

In addition to the two-point statistics above, Doppler lensing allows us to make spatially resolved maps of a quantity related to the gravitational potential. Combining equations (\ref{oi}) and (\ref{eq:kappav}) we find that 
\begin{equation}
\Phi'+a H \Phi=-\int d\chi \frac{3\Omega_{m}H_{0}^{2}}{2}\frac{\chi H}{\chi a H-1}\kappa_{v}(z,\bm n)
\end{equation}
Hence by combining our measured estimates of $\kappa_v (z,\bm n)$ with models or measurements of the other quantities in the integral, we are able to map the quantity on the right hand side. 

Alternatively, we could make a map of a quantity $\phi$ which is closer to the data, which we will call the Doppler lensing potential, being simply the sum of $\kappa_v$ along the line of sight, 
\begin{eqnarray}
\phi(z,\bm n)& =& \int_0^z dz' \kappa_{v}(z',\bm n)\\
& = &-\int_0^\chi d\chi' \left[\frac{a\chi' H-1}{ \chi'}\right] \frac{2}{3\Omega_{m}H_{0}^{2}}\partial_\chi(\Phi'+a H \Phi)\nonumber \\
\end{eqnarray}
which is a map combining information about geometry and potential; it is somewhat analogous to the lensing potential in gravitational lensing, which also includes an integral along the line of sight involving the gravitational potential and geometric factors. However, the two potentials behave differently: the gravitational lensing potential can be considered to be a 2-D projection of the gravitational potential, and varies only slowly with source redshift. On the other hand,  the Doppler lensing potential is an integral of the derivative of quantities including the gravitational potential, so varies rapidly with source redshift in a similar way to the gravitational potential itself. 

\begin{figure}
   \includegraphics[width=.5\textwidth]{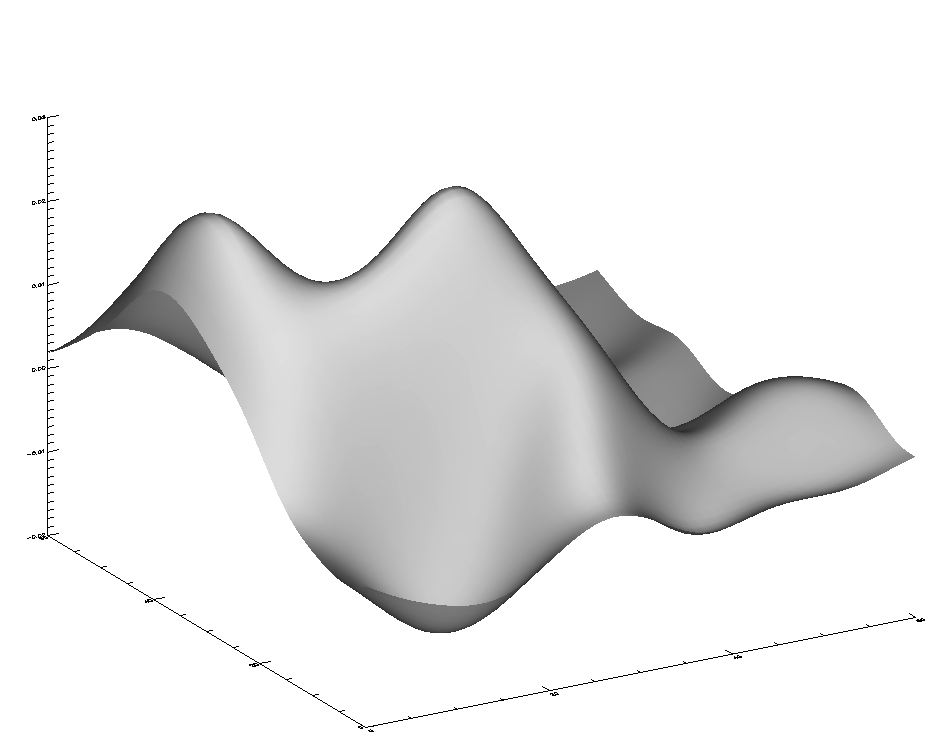} 
   \includegraphics[width=.5\textwidth]{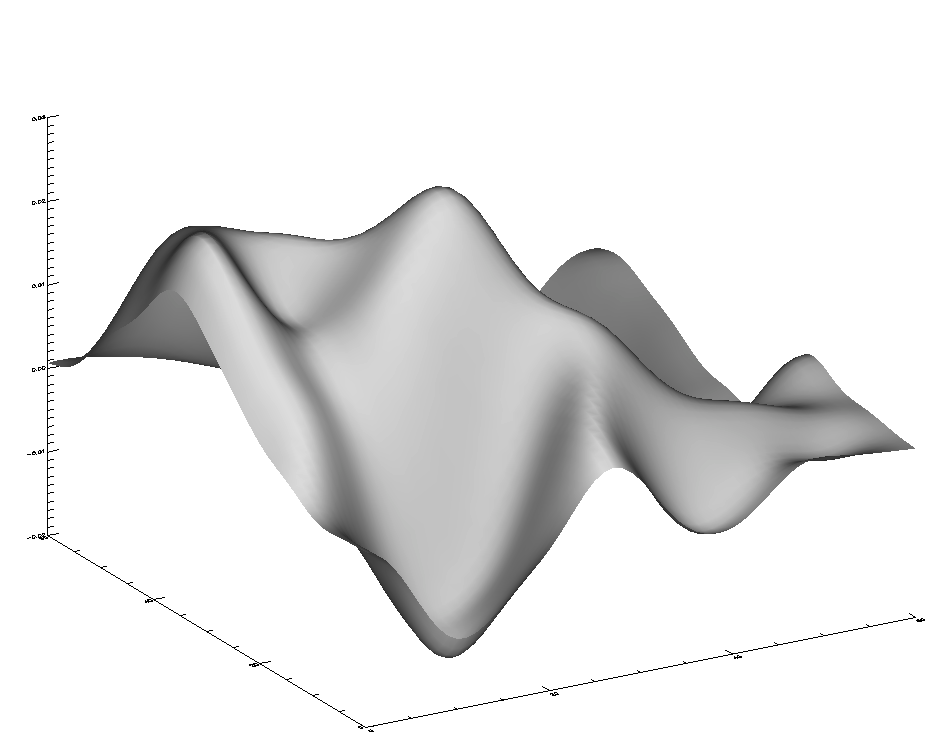} 
  \caption{Top panel: Smoothed Doppler potential map for a slice at $z$=0.2. Here, no noise has been added to the Doppler convergence at each galaxy, so this represents the true field we seek to reconstruct. Bottom panel: Smoothed Doppler potential map for the same slice, now with noise added to the Doppler convergence at each galaxy.}
  \label{fig:maprec}
\end{figure}

If we wish to consider the smoothed $\phi$ field, $\phi_s$ given by
\begin{equation}
\phi_s(z)=\int_0^\infty dz' \phi(z') W'(z'-z)
\end{equation}
where $W'$ is the convolving kernel, then integrating by parts and noting that $\kappa_v=\partial\phi/\partial z$, we find
\begin{equation}
\phi_s(z)=\int_0^\infty dz' \kappa_v(z') W(z'-z) 
\label{eq:phis}
\end{equation}
where $W$ is the integral with respect to redshift of $W'$. Here we choose
\begin{eqnarray}
W(z'-z)&=&(z'-z) \exp{\left(-\frac{(z'-z)^2}{2\sigma_z^2}\right)}, \label{eq:wz}\\
W'(z'-z)&=&\left(1-\frac{(z'-z)^2}{\sigma_z^2}\right)\exp{\left(-\frac{(z'-z)^2}{2\sigma_z^2}\right)}.
\end{eqnarray}
We make $\phi_s$ maps with our $50^\circ \times 50^\circ$ simulation (described in Section 3), for both the noise-free case and the case where $\kappa_v$ has realistic noise on each galaxy. We choose a smoothing radius of $\sigma_z=0.02$ for the radial convolving kernel above, and in addition smooth transversely with a Gaussian, with smoothing radius of $3^\circ$. An example slice at $z=0.2$ of the true $\phi_s$ field  is shown in the top panel of Figure \ref{fig:maprec}, together with the reconstructed $\phi_s$ field from noisy convergence data in the lower panel. We see that the reconstructed field bears a strong resemblance to the true $\phi_s$ field, with peaks and troughs in the field faithfully reproduced. 

\begin{figure}
   \includegraphics[width=.5\textwidth]{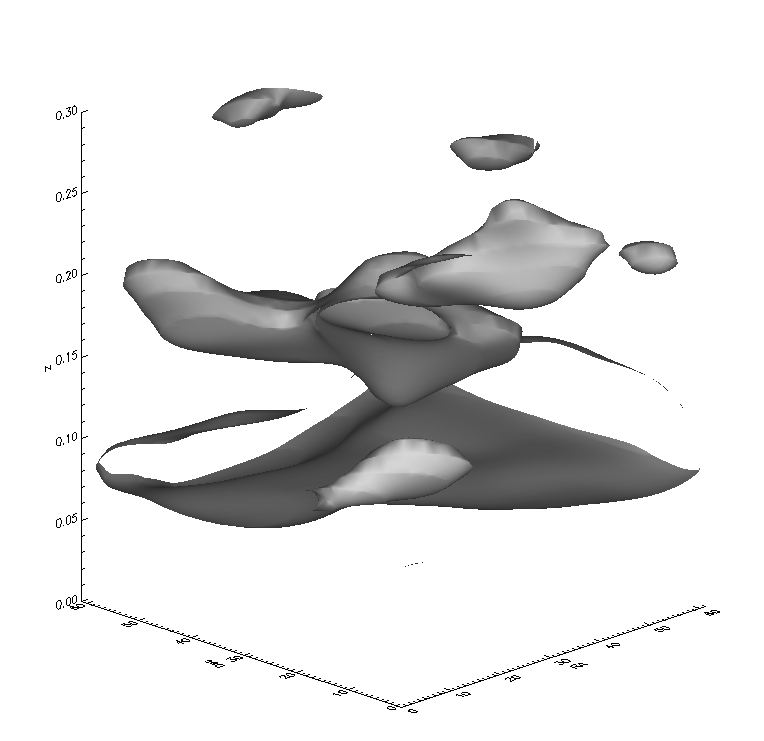} 
   \includegraphics[width=.5\textwidth]{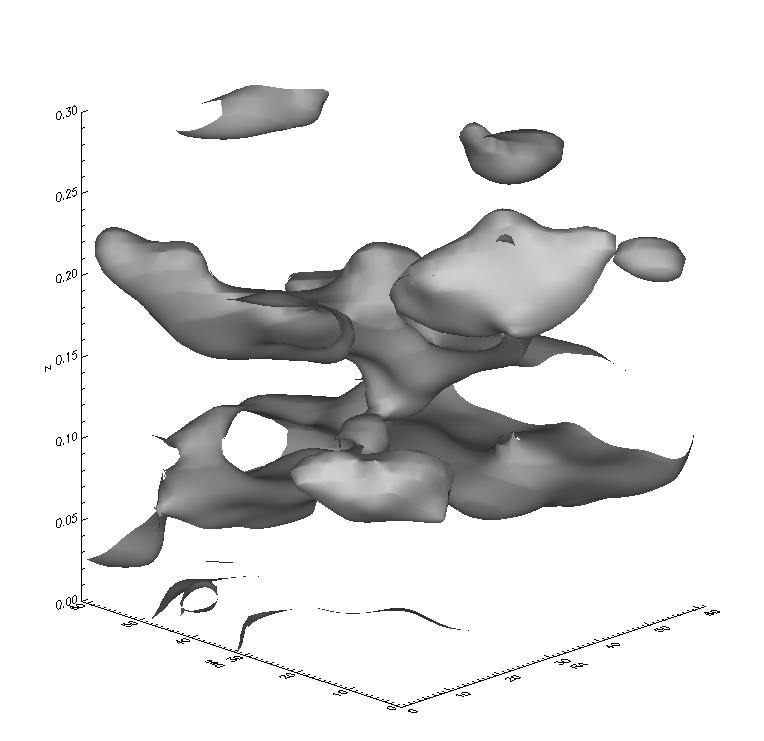} 
  \caption{Top panel: Smoothed Doppler potential 3D map, at isocontour $\phi_s=0.01$. Here, no noise has been added to the Doppler convergence at each galaxy. Bottom panel: Smoothed Doppler potential map for the same isocontour, now with realistic noise added to the Doppler convergence at each galaxy.}
  \label{fig:maprec3d}
\end{figure}

An alternative representation of the field is shown in Figure \ref{fig:maprec3d}; the top panel shows the true $\phi_s$ field in 3D, where an isocontour $\phi_s=0.01$ has been drawn. The lower panel shows the reconstructed $\phi_s$ field, in the presence of realistic convergence noise. Again we see that the Doppler lensing potential appears to be well estimated in this large 3D volume.

\begin{figure}
   \includegraphics[width=.48\textwidth]{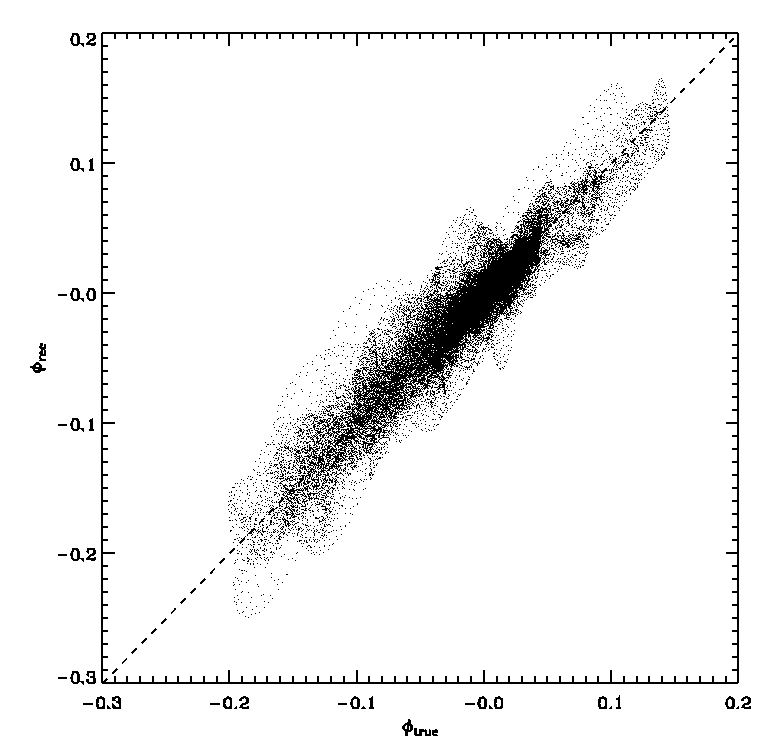}
  \caption{Reconstructed Doppler potential versus true Doppler potential for the wide field simulation, $0<z<0.3$.} 
  \label{fig:inout}
\end{figure}

This is quantified for the full 3-D field in Figure \ref{fig:inout}, which shows the Doppler lensing potential pixel values of the true versus reconstructed fields. The best fit line between these quantities is $\phi_{\rm rec}=0.96\phi_{\rm true}-3.8\times 10^{-4}$, and the Pearson correlation coefficient is 0.95 for $1.2\times 10^5$ pixels, which represents good agreement. 

\section{Conclusions}

In this paper we have investigated Doppler lensing as a probe of cosmology. This effect causes a slight change in the inferred physical size and brightness of objects at a given observed redshift; the magnitude of this effect is dependent on the peculiar velocities of galaxies, and so measurements of Doppler lensing are useful as a means of constraining the growth of structure and hence dark energy and gravitational physics.

As with gravitational lensing, the intrinsic range of sizes and magnitudes of galaxies leads to a substantial noise on Doppler lensing which needs to be overcome by using statistics which average over many galaxies. 

A difference between gravitational and Doppler lensing is their behaviour as a function of radial distance from a `lens'. Gravitational lensing convergence grows slowly behind a lens, whereas the amplitude of Doppler lensing rapidly rises then drops to zero both in front of and behind a lens (with different signs of the effect). Because of this, it is of great value to measure spectroscopic redshifts for sources when measuring Doppler lensing because of the rapid variation of the effect with redshift. 

We have explained the theoretical background for the effect in Section 2, showing how it originates in the alteration of redshift due to peculiar velocity, together with the fact that we infer distances based on observed redshifts. We showed that Doppler lensing dominates over gravitational lensing at medium-low redshifts and wavenumbers ($\ell \la 1000$ at $z=0.2$, and $\ell \la 100$ at $z=0.4$). 

We have examined the Doppler lensing effect in a series of simulations based on the Millennium simulation \citep{2005Natur.435..629S,2009MNRAS.398.1150B}. We have described these in Section 3, showing that they contain the expected Doppler lensing behaviour of having large convergence each side of a cluster or void, and of having a larger typical amplitude than gravitational lensing for $0<z<0.3$. We have  defined three survey configurations (Section 4) typical of forthcoming surveys, covering a fraction of the sky ranging from 0.12 to 0.75, and including dense spectroscopic redshift coverage for at least $0.1<z<0.3$.

We showed how one can measure the Doppler lensing convergence in spheres around stacked over- or under-densities (Section 5). We found that the signal is measurable with a stack of density peaks/voids, but is not detectable on individual objects. 

We then calculated two point statistics for the overdensity-Doppler convergence cross-correlation, and the Doppler convergence autocorrelation (Section 6). In each case, we correlate pairs which are close to each other in redshift ($\Delta z<0.02$) as it is on these scales that the Doppler lensing generates coherent convergence. 
We find that the two point correlations function can be measured with e.g. signal-to-noise $\simeq 80$ on $10'$ scales (survey 1), leading to useful constraints on cosmological parameters. We have discussed the potential systematic effects on the correlation functions due to intrinsic size correlations and gravitational lensing; for the cross-correlation we propose, we have shown that these systematics are small, while for the autocorrelation careful modelling of the intrinsic size correlation will be required.

Finally, we have shown how measurements of Doppler convergence can be used to make a 3D map of a potential field closely related to the gravitational potential. The reconstructed potential is strongly correlated with the true potential if smoothed on 50Mpc scales, despite the large noise term on convergence estimators.

In addition to the measurements proposed above, there are several further interesting areas for investigation with Doppler lensing. The effect gives us a direct way of estimating the peculiar velocity field at each galaxy, and therefore allows us to infer the velocity histogram for a volume of the Universe, and higher-point velocity statistics in addition to two-point statistics (redshift space distortions of the galaxy correlation function only provide the latter). Each of these quantities is sensitive to the evolution of large-scale structure, and can provide constraints on cosmology and gravitation.

In addition, one could use Doppler convergence averaging around particular selections of clusters or galaxy types in order to probe the typical velocity field and gravitational potential field around these objects. This is equivalent to the cluster lensing and galaxy-galaxy lensing approaches well known in weak gravitational lensing studies.

Finally, one could combine the auto- and cross-correlation function of Doppler convergence with auto- and cross- correlation functions for galaxy counts and gravitational lensing. Fitting theory jointly to all of these correlations should provide improved constraints on bias and growth parameters.

To conclude, Doppler lensing affords a novel range of cosmological applications. Forthcoming surveys, furnished with spectroscopic redshifts, will be able to make the most out of this exciting new cosmological probe.

\section*{Acknowledgements}

We thank Camille Bonvin, Sandro Ciarlariello, Rob Crittenden, Eric Huff, Peter Melchior, Bob Nichol, Lado Samushia, Rafal Szepietowski and Obinna Umeh for very useful discussions. DB and RM are supported by UK Science and Technology Facilities Council, grant ST/K00090X/1. SA and RM are supported by the South African Square Kilometre Array Project and the National Research Foundation, and a joint Royal Society (UK)/ NRF (SA) exchange grant. Please contact the authors to request access to research materials discussed in this paper. 

\appendix

\section{Gravitational lensing convergence}

The standard gravitational lensing term is given by the first line of equation \eqref{kphi}. The second line of \eqref{kphi} follows under the approximations $\nabla_\perp^2\approx\nabla^2$ and $\nabla^2\Phi = {3 H_0^2 \Omega_m }  \delta/(2a)$, which are based on the sub-Hubble limit of equations \eqref{spl} and \eqref{grp} respectively. 
With
 \be
 \bm n\cdot\bm \nabla=-\frac{d}{d\chi}-\frac{\partial}{\partial\eta},
 \ee 
equation \eqref{spl} implies
\be
\nabla_\perp^2\Phi=\nabla^2\Phi-{d\Phi\over d\chi^2}-2{d\Phi\over d\chi}-\Phi''+{2\over \chi}\left(\Phi'+{d\Phi\over d\chi}\right).
\ee
On sub-Hubble scales we can neglect all terms on the right after the first.
On these scales, the Poisson equation \eqref{grp} reduces to its Newtonian form, since we can neglect the $\Phi$ and $\Phi'$ terms. This then leads to the usual lensing magnification in the second line of equation \eqref{kphi} --  as an integral over the density contrast along a line of sight.

\section{Angular power spectra}
Here, we present full derivations of all the $C_{\ell}$ in Section 6. 

\subsection{Cross-correlation $C^{\delta \kappa_{v}}_{\ell}$}
 
In order to calculate the cross-power spectrum $C^{\delta \kappa_{v}}_{\ell}$, we need to obtain spherical harmonic decomposition coefficients for both the overdensity and the Doppler convergence.

In Fourier space, the potential can be decomposed as
\begin{equation}
\Phi(\bm x,\eta) = \int \frac{d^3\bm k}{(2\pi)^{3/2}}\Phi(\bm k,\eta)e^{i\bm k \cdot \bm x}
\end{equation}
and the power spectrum is defined by
\begin{equation}\label{a6}
 \langle \Phi(\bm k,\eta)\Phi^*(\bm k',\eta')\rangle=\frac{2\pi^2}{k^3}\mathcal{P}_{\Phi}(k,\eta,\eta')
 \delta^{(3)}(\bm k-\bm k')\,.\ 
\end{equation}
(We use the same convention for other variables.)
Taking the Fourier transform of equation \eqref{grp}, ignoring the sub-dominant term proportional to $(a\Phi)'$, and expanding the plane wave in spherical harmonics, the galaxy overdensity reads 
\ba
&&\delta_g(z,\bm \theta) = - \frac{2}{3H_{0}^{2}\Omega_{m}}a(z) b(z)\\&&
 \times4\pi\sum_{\ell m}\int \frac{d^3\bm k}{(2\pi)^{3/2}}k^{2}\Phi(\bm k,\eta) i^\ell j_\ell(k\chi)Y_{\ell m}(\hat{\bm k})Y_{\ell m}(\bm \theta)\,,\nonumber
\ea
where we have expanded the exponential in terms of spherical Bessel functions $j_\ell$ in the usual way. Here as in the text $\bm\theta=-\bm n$ is the direction of observation, opposite to the photon direction. In this equation we have introduced a linear bias $b(z)$ relating the galaxy overdensity to the total matter overdensity, $\delta_g(z)=b(z) \delta(z)$; from now on we will suppress the subscript $g$.

We now introduce a window function $W_1(z)$ over which we will consider the overdensity (this is the range of one tomographic bin). We can write the averaged overdensity in this thick redshift shell as
\ba
&&\delta(\bm \theta) = - \frac{8\pi}{3H_{0}^{2}\Omega_{m}}\sum_{\ell m}\int_{z_\text{min}}^{z_\text{max}}dzW_1(z)a(z)b(z)\\&&
 \times\int \frac{d^3\bm k}{(2\pi)^{3/2}}k^{2}\Phi(\bm k,\eta) i^\ell j_\ell(k\chi)Y_{\ell m}(\hat{\bm k})Y_{\ell m}(\bm \theta)\nonumber
\ea
We are then able to read off the coefficients for the spherical harmonic decomposition of $\delta(\bm \theta)$,
\ba
&&\delta_{\ell m}= -\frac{8\pi}{3H_{0}^{2}\Omega_{m}}\int_{z_\text{min}}^{z_\text{max}}dzW_1(z)a(z)b(z)\\&&
 \times\int \frac{d^3\bm k}{(2\pi)^{3/2}}k^{2}\Phi(\bm k,\eta) i^\ell j_\ell(k\chi)Y_{\ell m}(\hat{\bm k}).\nonumber
\ea
Now we turn to the Doppler convergence. This is related to peculiar velocity by
\begin{equation}
\kappa_{v}(z,\bm \theta) = -\left[1-\frac{1+z}{\chi H}\right] \bm v \cdot \bm \theta
\end{equation}
and smoothing in redshift space yields
\begin{equation}
\kappa_{v}(\bm \theta) = \int_{z_\text{min}}^{z_\text{max}} dz W_2(z,z')\kappa_{v}(z,\bm \theta)
\end{equation}
The weight function $W_2$ is specified in the text; it depends additionally upon the redshift $z'$ of the $\delta$ bin which we will correlate with.
 The Fourier decomposition of the convergence reads 
\ba
&&\kappa_{v}(\bm \theta) = \frac{2}{3\Omega_{m} H_{0}^2}\int_{z_\text{min}}^{z_\text{max}}dz W_2(z,z')a(z)\left[1-\frac{1}{\chi a H}\right]\nonumber\\&& \times \int \frac{d^3\bm k}{(2\pi)^{3/2}}\Phi(\bm k,\eta)\bm\theta\cdot\bm\nabla(e^{i\bm k \cdot \bm x})  
\ea
We expand the exponential in spherical harmonics while taking its spatial derivative along the line of sight:
\begin{equation}
\bm\theta\cdot\bm\nabla \left( \exp(i\bm k \cdot \bm x)\right)=4\pi\sum_{\ell m}i^\ell kj'_\ell(k\chi)Y_{\ell m}(\hat{\bm k})Y_{\ell m}(\bm \theta)
\end{equation}
where prime($'$) on the Bessel function here denotes its derivative with respect to the argument.  
We therefore obtain
\ba
&&\kappa_{v}(\bm \theta) = \frac{8\pi}{3\Omega_{m} H_{0}^2}\int_{z_\text{min}}^{z_\text{max}}dz W_2(z,z')a(z)\left[1-\frac{1}{\chi a H}\right]\nonumber \\&&\times \sum_{\ell m}\int \frac{d^3\bm k}{(2\pi)^{3/2}}\Phi(\bm k,\eta)i^\ell kj'_\ell(k\chi)Y_{\ell m}(\hat{\bm k})Y_{\ell m}(\bm \theta)
\ea
and we can now read off the coefficients for the spherical harmonic decomposition of Doppler convergence,
\ba
&&\kappa^{v}_{\ell m} = \frac{8\pi}{3\Omega_{m} H_{0}^2}\int_{z_\text{min}}^{z_{\max}}dz W_2(z,z')a(z)\left[1-\frac{1}{\chi aH}\right]\nonumber \\&& \times \int \frac{d^3\bm k}{(2\pi)^{3/2}}\Phi(\bm k,\eta)i^\ell  kj'_\ell(k\chi)Y_{\ell m}(\hat{\bm k}).
\label{eq:kapvell}
\ea
The angular cross-power spectrum for a particular tomographic bin can now be easily deduced, as
\ba \label{e1}
&&\langle \delta_{\ell m}\kappa_{\ell' m'}^{v^{*}}\rangle = \delta_{\ell \ell'}\delta_{mm'}\frac{16\pi}{(3H^{2}_{0}\Omega_{m})^{2}}\int_{0}^{\infty}dk k^{2} T^{2}(k)\mathcal{P}_{\Phi_{i}}(k)\nonumber\\&&\!\!\!\!\!\!\!\!\!\!\!\!\!
\times \int_{z_\text{min}}^{z_\text{max}}dz W_{1}(z)a(z)b(z)j_{\ell}(k\chi(z))g(z)\\&&\!\!\!\!\!\!\!\!\!\!\!\!\!
\times \int_{z_\text{min}}^{z_\text{max}}dz'W_{2}(z',z) \left[\frac{1}{\chi(z') H(z')}-a(z')\right]\mathcal{G}(z')j'_{\ell}(k\chi(z'))\nonumber 
\ea
where we have used equation \eqref{a6}. We have integrated over $\bm k'$, and used $d^3\bm k=k^2 dk d\hat k$; we have then integrated the product of spherical harmonics over $\hat{\bm k}$ to get $\delta_{\ell\ell'}\delta_{mm'}$. 

In addition, we have written the potential power spectrum $\mathcal{P}_{\Phi}$ in terms of the primordial power spectrum $\mathcal{P}_{\Phi_{i}}(k)$, 
\begin{equation}
\mathcal{P}_{\Phi}(k,\chi,\chi') = \mathcal{P}_{\Phi_i}(k)\tilde T(k,\chi)\tilde T(k,\chi')
\end{equation}
where
\be
 \tilde T(k,\chi) = T(k)\mathcal{G}(\chi)
\ee
and 
\be
\mathcal{G}(\chi) = g'(\chi)+a H g(\chi)
\label{eq:gcal}
\ee
where $g(\chi)$ is the growth suppression factor, and $T(k)$ is the transfer function. 

It is then straightforward to deduce $C^{\delta \kappa_{v}}_{\ell}$ for a particular tomographic bin from equations (\ref{eq:cdkav}) and (\ref{e1}):
\ba 
&&C^{\delta\kappa_{v}}_\ell= \frac{16\pi}{N\left(3H_{0}^{2}\Omega_{m}\right)^{2}}\int_{0}^{\infty} dk k^{2} T^{2}(k)\mathcal{P}_{\Phi_{i}}(k)\nonumber\\&&
\times \int_{z_\text{min}}^{z_\text{max}}dz W_{1}(z)g(z)a(z)b(z) j_{\ell}(k\chi(z)) \\&&
\times  \int_{z_\text{min}}^{z_\text{max}}dz' W_{2}(z',z) \left(\frac{1}{\chi H}-a\right)\mathcal{G}(z')j'_{\ell}(k\chi(z')).\nonumber
\ea

\subsection{Auto-correlation $C^{\kappa_{v}\kappa_{v}}_{\ell}$}

The Doppler lensing power spectrum $C^{\kappa_{v}\kappa_{v}}_{\ell}$ can be calculated in a very similar fashion. Using equation (\ref{eq:kapvell}) we obtain 
\ba
&&\langle \kappa_{\ell m}\kappa_{\ell' m'}^*\rangle= 4\pi\left(\frac{2}{3H_{0}^{2}\Omega_{m}}\right)^{2}\int_{0}^{\infty} dk kT^{2}(k)\mathcal{P}_{\Phi_{i}}(k)\label{e2}
\\&&\!\!\!\!\!\!\!\!\!\!\!\!\!
\times \int_{z_\text{min}}^{z_\text{max}}dz W_{1}(z) \left[a(z)-\frac{1}{\chi(z) H(z)}\right]\mathcal{G}(z)j'_{\ell}(k\chi(z))\nonumber\\&&\!\!\!\!\!\!\!\!\!\!\!\!\!
\times \int_{z_\text{min}}^{z_\text{max}}dz'W_{2}(z',z) \left[a(z')-\frac{1}{\chi(z')  H(z')}\right]\mathcal{G}(z')j'_{\ell}(k\chi(z'))\delta_{\ell \ell'}\delta_{mm'}\nonumber
\ea
where we have again used equation \ref{a6}, integrated over $\bm k'$, used $d^3\bm k=k^2 dk d\hat k$ and integrated the product of spherical harmonics over $\hat{\bm k}$ to get
$\delta_{\ell\ell'}\delta_{mm'}$. The fact that we use two different window functions ($W_{1}$, $W_{2}$) is explained in Section~6. 

Now it is straightforward to use equation (\ref{e2}) with equation (\ref{eq:ckkav}) to obtain the form of 
$C^{\kappa_{v}\kappa_{v}}$:
\ba
&&C^{\kappa_{v}\kappa_{v}}_\ell= \frac{16\pi}{N\left(3H_{0}^{2}\Omega_{m}\right)^{2}}\int_{0}^{\infty} dk kT^{2}(k)
\mathcal{P}_{\Phi_{i}}(k)\\&&
\times \left[\int_{z_\text{min}}^{z_\text{max}}dz W_{1}(z)\left(a-\frac{1}{\chi H}\right)\mathcal{G}(z)j'_{\ell}(k\chi(z))\right]\nonumber\\&& 
\times \left[\int_{z_\text{min}}^{z_\text{max}}dz'W_{2}(z',z) \left(a-\frac{1}{\chi  H}\right)\mathcal{G}(z')j'_{\ell}(k\chi(z'))\right]\nonumber
\ea

\bibliographystyle{mn2e}
\bibliography{bibliography}

\begin{thebibliography}{18}
\expandafter\ifx\csname natexlab\endcsname\relax\def\natexlab#1{#1}\fi

\bibitem[{{Bartelmann} \& {Schneider}(2001)}]{2001PhR...340..291B}
{Bartelmann} M., {Schneider} P., 2001, \physrep, 340, 291

\bibitem[{{Bolejko} {et~al}\mbox{.}(2013){Bolejko}, {Clarkson}, {Maartens},
  {Bacon}, {Meures}, \& {Beynon}}]{2013PhRvL.110b1302B}
{Bolejko} K., {Clarkson} C., {Maartens} R., {Bacon} D., {Meures} N., {Beynon}
  E., 2013, Physical Review Letters, 110, 021302

\bibitem[{{Bonvin}(2008)}]{2008PhRvD..78l3530B}
{Bonvin} C., 2008, \prd, 78, 123530

\bibitem[{{Bonvin} {et~al}\mbox{.}(2006){Bonvin}, {Durrer}, \&
  {Gasparini}}]{2006PhRvD..73b3523B}
{Bonvin} C., {Durrer} R., {Gasparini} M.~A., 2006, \prd, 73, 023523

\bibitem[{{Boylan-Kolchin} {et~al}\mbox{.}(2009){Boylan-Kolchin}, {Springel},
  {White}, {Jenkins}, \& {Lemson}}]{2009MNRAS.398.1150B}
{Boylan-Kolchin} M., {Springel} V., {White} S.~D.~M., {Jenkins} A., {Lemson}
  G., 2009, \mnras, 398, 1150

\bibitem[{{Dewdney} {et~al}\mbox{.}(2009){Dewdney}, {Hall}, {Schilizzi}, \&
  {Lazio}}]{2009IEEEP..97.1482D}
{Dewdney} P.~E., {Hall} P.~J., {Schilizzi} R.~T., {Lazio} T.~J.~L.~W., 2009,
  IEEE Proceedings, 97, 1482

\bibitem[{{Eisenstein} \& {Hu}(1998)}]{1998ApJ...496..605E}
{Eisenstein} D.~J., {Hu} W., 1998, \apj, 496, 605

\bibitem[{{Guo} {et~al}\mbox{.}(2010){Guo}, {White}, {Li}, \&
  {Boylan-Kolchin}}]{2010MNRAS.404.1111G}
{Guo} Q., {White} S., {Li} C., {Boylan-Kolchin} M., 2010, \mnras, 404, 1111

\bibitem[{{Heavens} {et~al}\mbox{.}(2013){Heavens}, {Alsing}, \&
  {Jaffe}}]{2013MNRAS.433L...6H}
{Heavens} A., {Alsing} J., {Jaffe} A.~H., 2013, \mnras, 433, L6

\bibitem[{Howlett {et~al}\mbox{.}(2012)Howlett, Lewis, Hall, \&
  Challinor}]{Howlett:2012mh}
Howlett C., Lewis A., Hall A., Challinor A., 2012, JCAP, 1204, 027

\bibitem[{{Hu} \& {Jain}(2004)}]{2004PhRvD..70d3009H}
{Hu} W., {Jain} B., 2004, \prd, 70, 043009

\bibitem[{{Joachimi} {et~al}\mbox{.}(2013){Joachimi}, {Semboloni}, {Hilbert},
  {Bett}, {Hartlap}, {Hoekstra}, \& {Schneider}}]{2013MNRAS.436..819J}
{Joachimi} B., {Semboloni} E., {Hilbert} S., {Bett} P.~E., {Hartlap} J.,
  {Hoekstra} H., {Schneider} P., 2013, \mnras, 436, 819

\bibitem[{{Kaiser} {et~al}\mbox{.}(1995){Kaiser}, {Squires}, \&
  {Broadhurst}}]{1995ApJ...449..460K}
{Kaiser} N., {Squires} G., {Broadhurst} T., 1995, \apj, 449, 460

\bibitem[{{Laureijs} {et~al}\mbox{.}(2011){Laureijs}, {Amiaux}, {Arduini},
  {Augu{\`e}res}, {Brinchmann}, {Cole}, {Cropper}, {Dabin}, {Duvet}, {Ealet},
  \& et~al.}]{2011arXiv1110.3193L}
{Laureijs} R. {et~al.}, 2011, ArXiv e-prints

\bibitem[{Lewis {et~al}\mbox{.}(2000)Lewis, Challinor, \&
  Lasenby}]{Lewis:1999bs}
Lewis A., Challinor A., Lasenby A., 2000, Astrophys. J., 538, 473

\bibitem[{{Schmidt} {et~al}\mbox{.}(2012){Schmidt}, {Leauthaud}, {Massey},
  {Rhodes}, {George}, {Koekemoer}, {Finoguenov}, \&
  {Tanaka}}]{2012ApJ...744L..22S}
{Schmidt} F., {Leauthaud} A., {Massey} R., {Rhodes} J., {George} M.~R.,
  {Koekemoer} A.~M., {Finoguenov} A., {Tanaka} M., 2012, \apjl, 744, L22

\bibitem[{{Smith} {et~al}\mbox{.}(2003){Smith}, {Peacock}, {Jenkins}, {White},
  {Frenk}, {Pearce}, {Thomas}, {Efstathiou}, \&
  {Couchman}}]{2003MNRAS.341.1311S}
{Smith} R.~E. {et~al.}, 2003, \mnras, 341, 1311

\bibitem[{{Springel} {et~al}\mbox{.}(2005){Springel}, {White}, {Jenkins},
  {Frenk}, {Yoshida}, {Gao}, {Navarro}, {Thacker}, {Croton}, {Helly},
  {Peacock}, {Cole}, {Thomas}, {Couchman}, {Evrard}, {Colberg}, \&
  {Pearce}}]{2005Natur.435..629S}
{Springel} V. {et~al.}, 2005, \nat, 435, 629

\end{thebibliography}

\label{lastpage}

\end{document}